\documentclass[traditabstract]{aa}

\usepackage[varg]{txfonts}

\usepackage{natbib}
\bibpunct{(}{)}{;}{a}{}{,} 

\usepackage{xspace}
\usepackage{graphicx}
\usepackage{xfrac}
\usepackage{amsmath}
\usepackage[colorlinks=true, urlcolor=blue, linkcolor=blue, citecolor=blue]{hyperref}

\newcommand{\Planck}{{\it Planck}\xspace}

\newcommand{\FlexKnot}{\texttt{FlexKnot}\xspace}

\providecommand{\sorthelp}[1]{}

\graphicspath{{plots/}{/}}

\begin{document}

\title{Cosmic Microwave Background Constraints in Light of Priors Over Reionization Histories}

\author{Marius Millea\inst{\ref{ilp},\ref{iap},\ref{email}} \and Fran{\c c}ois Bouchet\inst{\ref{iap}}}

\institute{ Institut Lagrange de Paris (ILP), Sorbonne Universit\'es, 98bis boulevard Arago, F-75014 Paris, France\label{ilp} \and Institut d'Astrophysique de Paris (IAP), UMR7095, CNRS \& Sorbonne Universit\'es, 98bis boulevard Arago, F-75014 Paris, France\label{iap} \and \email{mariusmillea@gmail.com} \label{email}}

\abstract{Non-parametric reconstruction or marginalization over the history of reionization using cosmic microwave background data necessarily assumes a prior over possible histories. We show that different but reasonable choices of priors can shift current and future constraints on the reionization optical depth, $\tau$, or correlated parameters such as the neutrino mass sum, $\Sigma m_\nu$, at the level of 0.3--0.4\,$\sigma$, i.e., that this analysis is somewhat prior dependent. We point out some prior-related problems with the commonly used principal component reconstruction, concluding that the significance of some recent hints of early reionization in \Planck 2015 data has been overestimated. We also present the first non-parametric reconstruction applied to newer \Planck intermediate (2016) data and find that the hints of early reionization disappear entirely in this more precise dataset. These results limit possible explanations of the EDGES 21cm signal which would have also significantly reionized the universe at $z\,{>}\,15$. Our findings about the dependence on priors motivate the pursuit of improved data or searches for physical reionization models which can reduce the prior volume. The discussion here of priors is of general applicability to other non-parametric reconstructions, for example of the primordial power spectrum, of the recombination history, or of the expansion rate.}


\maketitle

\section{Introduction} 
\label{sec:intro}

The exact details of the ``epoch of reionization'', during which the universe transitioned from a mostly neutral hydrogen gas into the highly ionized state we see today, are still of considerable uncertainty. This transition left several imprints on the Cosmic Microwave Background (CMB) which can be used to constrain the physics of reionization, but also serve as a ``nuisance'' uncertainty which must be marginalized over when considering constraints on other parameters. 

One of the most widely used methods for reconstructing or marginalizing over the reionization history from CMB data has been based on decomposing the history into eigenmodes \citep[][hereafter HH03]{hu2003}. This has been applied to a number of datasets, for example the WMAP data \citep[][hereafter MH08]{mortonson2008a}, and most recently the \Planck 2015 data \citep{heinrich2017,heinrich2018,obied2018}. The latter works argue that the generic reconstruction reveals a $>\!2\,\sigma$ preference for a high-redshift ($z\gtrsim 15$) contribution to the optical depth, $\tau$. Conversely, other works have used alternate methods and not found such evidence in this same dataset \citep{hazra2017,villanueva-domingo2018}. 

In this work, we will point out two problems with PCA analyses that have to-date been largely or completely overlooked. Correcting these problems will lead to finding reduced evidence of early reionization in the \Planck 2015 data, more consistent with the latter results. 

The first problem has been mentioned in \cite{mortonson2008a}, \cite{heinrich2018}, and described in some detail in \cite{lewis2006}, although not explicitly in relation to the PCA model. The issue is that taking a flat prior on amplitudes of the reionization principal components induces a non-flat prior on $\tau$. \cite{heinrich2018} argue that this effect is unimportant, but do not perform direct or fully conclusive tests. Here we perform the very direct test of simply calculating the effective $\tau$ prior induced by the flat mode priors, finding that it is roughly $\tau\approx0.20\,{\pm}\,0.07$. Using a maximum entropy procedure, we remove the effects of this prior, finding that it accounts for a significant part of the shift to higher $\tau$ reported by \cite{heinrich2017}.

Another problem with existing PCA analyses lies in the choice of physicality priors. These priors are necessary to ensure that the reconstructed ionization fraction at any given redshift remains within physical range (i.e., remains positive but less that the maximum corresponding to a fully ionized universe). Due to the nature of the PCA decomposition, the physicality priors necessarily allow {\it some} unphysical models (this may seem quite counter-intuitive, but we will give a simple geometric picture of why this is indeed the case). However, existing analyses have used a sub-optimal set of physicality priors which allowed {\it more} unphysical models than necessary. We will derive the optimal set of priors, and show that switching to these leads to significant changes in constraints. Together with removing the impact of the informative $\tau$ prior, we find that evidence for early reionization in the \Planck 2015 data is reduced from $>\!2\,\sigma$ to $1.4\,\sigma$. 

Since the PCA results depend significantly on the details of the physicality priors, which will always be necessarily imperfect, we conclude that the PCA procedure is poorly tailored to the problem of generic reionization history reconstruction. We will seek a better method, in particular one which does not allow unphysical models. To this end, we propose what we will call the \FlexKnot model. This model interpolates the history using a varying number of knots, with the exact number of knots determined by the data itself via a Bayesian evidence calculation. The model is somewhat similar to the \cite{hazra2017} approach, but improves upon it by not requiring an a posteriori and fixed choice of knot positions, which otherwise creates undesired regions of high prior probability at certain redshifts. The \FlexKnot model follows almost exactly the same procedure used previously to generically reconstruct the primordial power spectrum from \Planck data \citep{vazquez2012,planck2014-a24}, and is particularly well suited to the problem here as well.

Using these reionization models, we calculate new constraints on the history of reionization coming from \Planck intermediate data, in particular using the \texttt{simlow} likelihood \citep{planck2014-a10}. We will refer to the combination of \Planck 2015 data and the \texttt{simlow} likelihood as the \Planck 2016 data. This likelihood has already been used to explore various parametric reionization models \citep{planck2014-a25}; however, a generic reconstruction fully capable of detecting an early component to reionization (should it be there) has not yet been performed. We report results from this particular dataset here.

The \Planck 2016 data represents a significant increase in constraining power on reionization as compared to the 2015 data, and we find that the remaining $1.4\,\sigma$ hints of an early reionization component are completely erased. Instead, we tightly constrain the $z\,{>}\,15$ contribution to the optical depth to be $<0.015$ at 95\% confidence, and find good agreement with a late and fast transition to a fully ionized universe, consistent with the conclusions of \cite{planck2014-a25}.

Even though the \texttt{simlow} data are quite constraining, some dependence on priors remains when performing generic reconstruction. In particular, the difference between a flat prior on $\tau$ and a flat prior on the knot positions leads to an $0.4\,\sigma$ shift in the resulting $\tau$ constraint. Both of these priors are fairly reasonable, and we consider it difficult to argue very strongly for one or the either of these priors in a completely generic setting. We thus consider this indicative of a level of systematic uncertainty stemming from our imperfect knowledge of the model, which we then propagate into shifts on other parameters which are correlated with $\tau$, given both current data and future forecasts.

We specifically focus on the sum of the neutrino masses, $\Sigma m_\nu$, which is expected to be measured extremely precisely by next generation CMB experiments \citep{abazajian2015}. Using Fisher forecasts, we show that a possible $0.4\,\sigma$ bias on $\tau$ from switching priors manifests itself as a $0.3\,\sigma$ bias on $\Sigma m_\nu$ given expect constraints from a ``Stage 4'' CMB experiment \citep[CMB-S4;][]{abazajian2016a}. We comment on the ability of combining with other probes, such as DESI-BAO, or a cosmic-variance-limited CMB large-scale polarization measurement, to reduce this uncertainty. 

Given these biases, we conclude that generic reconstructions of the reionization history are likely not enough to achieve the most accurate bounds on $\Sigma m_\nu$. This motivates work on obtaining constraints with {\it physical} models which have free parameters that can realistically account for our lack of exact knowledge about the physics of reionization. Although not directly used throughout the paper, will also comment on the possibility for other probes of reionization to reduce the dependence on priors, specifically measurements of the patchy kinetic Sunyaev-Zeldovich effect \citep{gruzinov1998,knox1998} and direct probes of the ionization state of the inter-galactic medium \citep{bouwens2015}.

The paper is outlined as follows. In Sec.~\ref{sec:models} we discuss the reionization models we consider, and in Sec.~\ref{sec:implicit_tau} we focus on the implicit $\tau$ prior induced by these procedures. These sections are a fairly technical description of the methodology used, and those wishing to see the results should skip to Secs.~\ref{sec:bflike}, \ref{sec:simlow}, and \ref{sec:future} where we discuss constraints from \Planck 2015, intermediate, and future data, respectively.


\section{Reionization models}
\label{sec:models}

\subsection{The TANH model}

The reionization history often assumed in CMB analyses because it has a useful parametric form and matches physical expectations somewhat well involves a single smooth step from an almost fully neutral universe\footnote{There is a small residual ionization level on the order of $10^{-4}$ remaining after recombination. We ignore this as the data considered here is insensitive to it.} to one with hydrogen fully ionized and helium singly ionized. The free electron fraction, in our convention the ratio between the number density of free electrons and hydrogen nuclei, $x_{\rm e} \equiv n_{\rm e}/n_{\rm H} $, is taken to be
\begin{align} \label{eq:tanh}
    x_{\rm e}^{\rm TANH}(z) =  \frac{1+f_{\rm He}}{2}\left\{1+\tanh\left[\frac{y(z^*) - y(z)}{\Delta y}\right]\right\},
\end{align}
with $y(z) = (1+z)^{3/2}$ and $\Delta y=3/2(1+z)^{1/2}\Delta z$, giving a transition centered at redshift $z^*$ with width $\Delta z\,{=}\,0.5$. Here the factor $f_{\rm He}$ is the number density ratio of helium to hydrogen nuclei (we have neglected all other atoms). We will refer to this model as the ``TANH'' model.

The contribution to the optical depth between any two redshifts $z_1$ and $z_2$ for this (or any) reionization history can be written as
\begin{align}
    \tau(z_1, z_2) = \sigma_T \int_{z_1}^{z_2} dz \,\frac{n_{\rm H}(z)(1+z)^2}{H(z)} x_{\rm e}(z),
\end{align}
where $\sigma_{\rm T}$ is the Thompson scattering cross-section. This is often used to parametrize Eq.~\eqref{eq:tanh} in terms of the total optical depth $\tau\equiv\tau(0,z_{\rm early})$ rather than $z^*$ (where $z_{\rm early}$ is some redshift before reionization began, but after recombination ended).

Note that in this convention for $x_{\rm e}$, the maximum ionization fraction can be greater than one, in particular can be as large as $x_{\rm e}^{\rm max}\equiv 1+f_{\rm He}$ before the second recombination of helium. We also note that second helium recombination is expected to be a small contribution to $\tau$, on the order of $\approx0.001$ depending on the exact values of other cosmological parameters. As this is already fairly small, we ignore any model-dependence in helium second reionization and model it as another transition with the same hyperbolic tangent form as in Eq.~\eqref{eq:tanh} but with $z_*\,{=}\,3.5$ and $\Delta z\,{=}\,0.5$, normalized such that it increases the ionization fraction from $x_{\rm e}^{\rm max}$ to $1+2f_{\rm He}$.

\subsection{The PCA model}
\label{sec:modes}

The TANH model has {\it some} parametric freedom, more-so if $\Delta z$ is also taken as a free parameter. However, we would like a model which can reproduce any arbitrary history, thus allowing us to generically reconstruct $x_{\rm e}(z)$ from the data. 

HH03 proposes a model based on decomposing the free electron fraction history into eigenmodes such that
\begin{align} \label{eq:hh03}
    x_{\rm e}^{\rm HH03}(z) = x_{\rm e}^{\rm fid}(z) + \sum_i m_i S_i(z)
\end{align}
for some fiducial $x_{\rm e}^{\rm fid}(z)$, eigenmode amplitudes $m_i$, and eigenmode templates $S_i(z)$. These templates have support between some $z_{\rm min}$ and $z_{\rm max}$ and are uniquely defined by three properties:
\begin{enumerate}
\item they form a special orthonormal basis\footnote{For clarity, we suppress writing the integration limits $z_{\rm min}$ and $z_{\rm max}$ for the rest of the paper. The exact normalization is purely a definitional choice. Similarly, the distinction of {\it special} orthonormal basis, i.e., that $S$ is purely a rotation, is not usually made, but we choose to do so here for definiteness and for aiding the subsequent geometrical interpretation.} for $x_{\rm e}(z)$, implying that $\int dz \,S_i(z) S_j(z) = \delta_{ij}$.
\item they diagonalize the covariance of the amplitude parameters given cosmic-variance-limited large-scale polarization data, $\langle \Delta m_i \Delta m_j \rangle=\sigma_i^2 \delta_{ij}$;
\item they are ordered by increasing $\sigma_i$.
\end{enumerate}

The lack of Gunn-Peterson absorption in quasars out to $z\approx6$ strongly constrains the universe to be very close to fully reionized by this redshift \citep{becker2001,fan2002}. We implicitly impose these bounds here by setting $z_{\rm min}\,{=}\,6$. Similarly as in other PCA analyses, we take $z_{\rm max}\,{=}\,30$.

\subsubsection{Geometric View of Physicality Priors}

\begin{figure*}

\includegraphics[width=\textwidth]{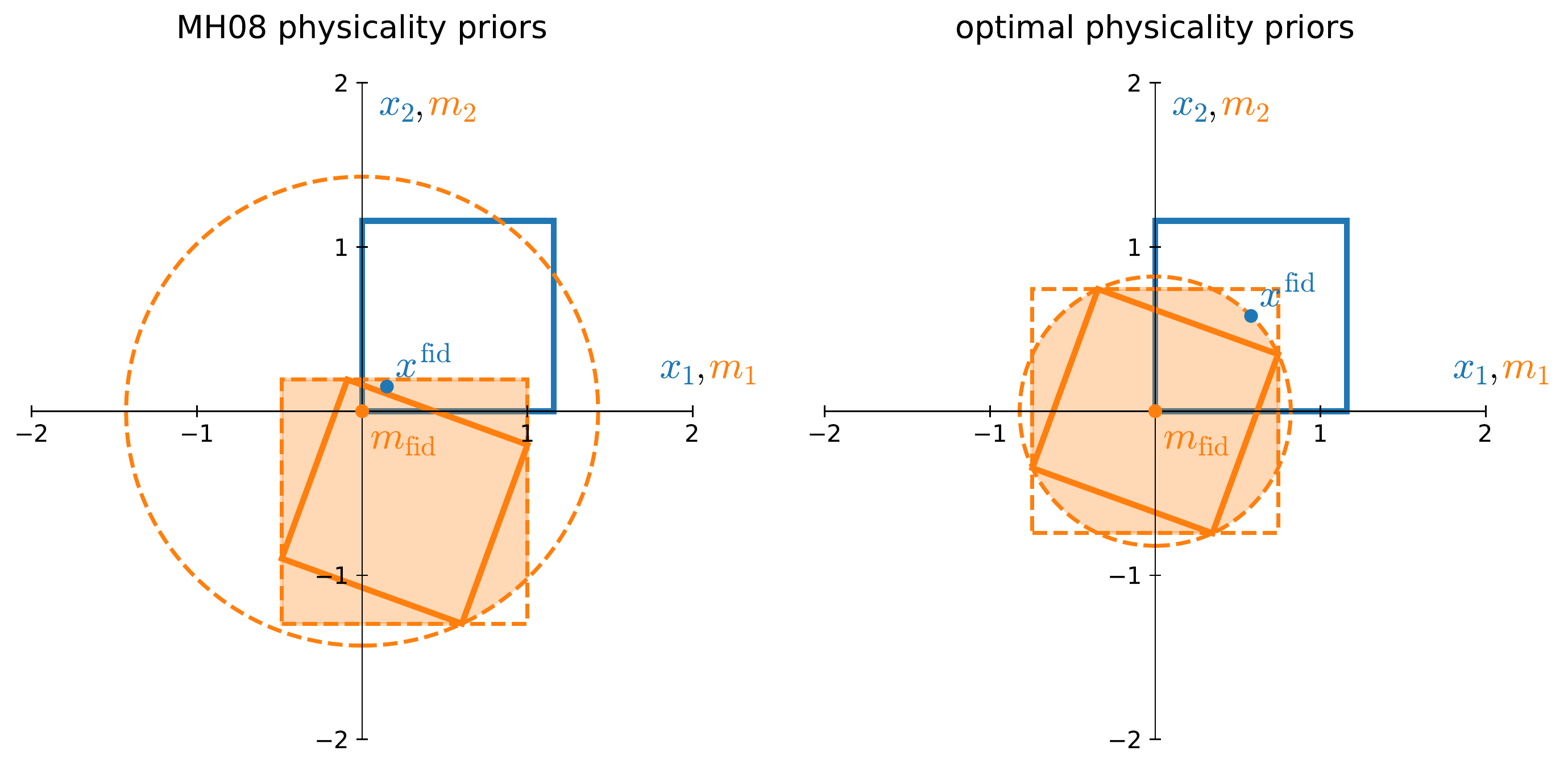}
\caption{A geometric picture of the PCA physicality priors in two-dimensions, demonstrating why they necessarily allow unphysical regions of parameter space. The solid blue square is the physicality region in terms of the ionization fraction, $x_1$ and $x_2$, at individual redshifts. The solid orange square is the physicality region in terms of the mode amplitudes, $m_1$ and $m_2$. The two types of physicality priors corresponding to Eq.~\eqref{eq:hyperbox_prior} and Eq.~\eqref{eq:hypersphere_prior} are shown as the dashed square and dashed circle, respectively. Their intersection, shaded orange, is the full physicality region. The left panel takes a fiducial ionization history, $x^{\rm fid}=0.15$, consistent with \cite{mortonson2008} and subsequent works. The right panel shows that the undesired but allowed unphysical parameter space can be reduced (but not fully) with a better choice of fiducial model.}
\label{fig:hyperbox}
    
\end{figure*}

The HH03 procedure also calls for a set of ``physicality priors'' on these mode amplitudes. These are necessary because otherwise nothing prevents the mode amplitudes from taking on values such that the bound $0<x_{\rm e}(z)<x_{\rm e}^{\rm max}$ fails to hold for some $z$, and this would have no physical meaning. \cite{mortonson2008} derive two sets of priors to enforce physicality. The first is an upper and lower bound on each $m_i$ such that $m_i^-\,{\le}\,m_i\,{\le}\,m_i^+$ with
\begin{align} \label{eq:hyperbox_prior}
    m_i^{\pm} = \int dz \left\{S_i(z)\left[1-2x_{\rm e}^{\rm fid}(z)\right] \pm \left|S_i(z)\right|\right\}/2.
\end{align}
The second is an upper bound on the sum of the squares of the mode amplitudes,
\begin{align} \label{eq:hypersphere_prior}
    \int dz \sum_i m_i^2  < \int dz \, \big[x_{\rm e}^{\rm max} - x_{\rm e}^{\rm fid}(z)\big]^2.
\end{align}

These priors may seem a bit abstruse, but they have a quite simple geometric interpretation which has not been highlighted before. We will give this interpretation here, aided by Fig.~\ref{fig:hyperbox}. This will also aid in spotting some improvements that can be made. 

To start, consider the reionization history, $x_{\rm e}(z)$, evaluated at $N$ redshifts, $z_i$, and stacked into a vector $x_i\equiv x_{\rm e}(z_i)$.  The physicality region in the $x_i$ vector space is trivial, each $x_i$ must individually fall between $0$ and $x_{\rm e}^{\rm max}$. This region is thus an $N$-dimensional hypercube with one vertex at the origin, extending into the positive hyper-quadrant, and with edge length $x_{\rm e}^{\rm max}$. 

The decomposition into mode amplitudes is given by applying the transformation $S$ to the residual between $x_i$ and the fiducial model, 
\begin{align} \label{eq:transform}
    m_i = \sum_j S_{ij}(x_j - x^{\rm fid}_j),
\end{align}
where $S_{ij}\equiv S_i(z_j)$, i.e., it is a matrix for which each row is one of the eigenmodes. The physicality region in the $m_i$ vector space is thus a translated and transformed version of the original hypercube. Note that because $S$ is special orthonormal and thus can only contain rotations, the region remains a hypercube.

Let us now visualize this transformation in two dimensions, represented in the left panel of Fig.~\ref{fig:hyperbox}. The solid blue square is the physicality region in terms of $x_i$, here just $x_1$ and $x_2$. The solid orange square is the same region after transformation by Eq.~\eqref{eq:transform}, assuming some arbitrary $S$ for demonstration purposes, and taking $x^{\rm fid}_j=0.15$ following \cite{mortonson2008} and all other subsequent PCA analyses.

If we were simultaneously fitting $m_1$ and $m_2$ in our analysis, the physicality region would be trivial to enforce, and would be given simply by the solid orange square. However, the point of the PCA procedure is that we do not need to simultaneously fit both parameters, as higher order modes like $m_2$ can be effectively marginalized over by just fixing them, since they have no observable impact and are decorrelated with the lower order modes like $m_1$. Usually we fix $m_2\,{=}\,0$, but any other value should work just as well. In designing the physicality prior for $m_1$, we need to ensure that all values that $m_1$ could take are allowed, not just those allowed when $m_2\,{=}\,0$, since this was just an artificial method of marginalization. Put another way, one must consider that a model which appears to be unphysical could be brought back into physicality by adjusting the higher order mode amplitudes which were artificially set to zero. This leads to two priors which can be constructed. 

The first prior is simply the bounding box of the solid orange square and is depicted as the dashed orange square. We will refer to this as the ``hyperbox'' prior\footnote{Note that in dimensions higher than the two depicted in Fig.~\ref{fig:hyperbox}, this region does not necessarily have equal edge length in all directions, and is thus truly a hyperbox as opposed to a hypercube.}. This prior just takes a hard bound on each $m_i$ corresponding to the largest and smallest possible value for that $m_i$ anywhere within the physicality region. This is indeed exactly what is calculated by Eq.~\eqref{eq:hyperbox_prior}. One can see this by noting that the $x_j$ which maximizes $m_i$ in Eq.~\eqref{eq:transform} is such that $x_j\,{=}\,x^e_{\rm max}$ if $S_{ij}$ is positive, and zero otherwise, and that Eq.~\eqref{eq:hyperbox_prior} is the transform of this $x_j$.

The second prior comes from noting that as long as $x^{\rm fid}\,{\leq}\,x_{\rm e}^{\rm max}/2$, the top right corner of the physicality region in terms of $x$ (i.e. the top right corner of the solid blue square in Fig.~\ref{fig:hyperbox}), will always be the furthest point from the origin even after transformation. Its distance from the origin will be unaffected by the rotation $S$, only by the translation by $x_{\rm fid}$, thus we can exclude points further than $\|x_{\rm e}^{\rm max} - x^{\rm fid}\|$ in the mode parameter space. This leads to the physicality region depicted by the dashed circle, and we will refer to this as the ``hypersphere'' prior. One can recognize this as the second MH08 prior in Eq.~\eqref{eq:hypersphere_prior} by noting that integral over $z$ there is analogous to the sum that appears in the vector norm, $\|x_{\rm e}^{\rm max} - x^{\rm fid}\|$.

The intersection of these two priors is shaded orange, and corresponds to the full MH08 physicality region. Note that although the two priors do remove some of the allowed unphysical regions of parameter space, they can never remove all of them. As we will see in Sec.~\ref{sec:simlow}, this unavoidably allowed unphysical region has undesirable consequences. 

While the amount of unphysical parameter space allowed by the hyperbox prior does not depend on our choice of $x^{\rm fid}$, this is not the case for the hypersphere prior. Indeed, the geometric picture makes it clear that we could reduce the allowed unphysical regions by picking $x^{\rm fid}\,{=}\,x^{\rm max}/2$, i.e., placing the fiducial model at the center of the original hypercube.\footnote{Equivalently, we could modify the hypersphere prior to be centered on the physicality region rather than on the origin, but we choose to discuss this in terms of picking a different $x^{\rm fid}$ simply for convenience so that Eq.~\eqref{eq:hypersphere_prior} is unchanged.} This is depicted in the right panel of Fig.~\ref{fig:hyperbox}. This choice is optimal in the sense that no other choice of $x^{\rm fid}$ excludes as much of the unphysical region, given the two priors we have constructed. As we will see in Sec.~\ref{sec:implicit_tau}, using the optimal physicality region is also useful because the increased symmetries in this case allow an analytic calculation of the induced $\tau$ prior. 

The physicality prior obtained by taking $x^{\rm fid}$ to be in the center of the physicality region is objectively better than other choices, and should be used. It does not bias the analysis in any other way, in fact it prevents biases from an over-allowance of unphysical parameters. We will show in Secs.~\ref{sec:bflike} and \ref{sec:simlow} that the ``sub-optimality'' of the MH08 prior has a significant impact on constraints from data.

\subsection{The HS17 model}

\begin{figure*}
\includegraphics[width=\textwidth]{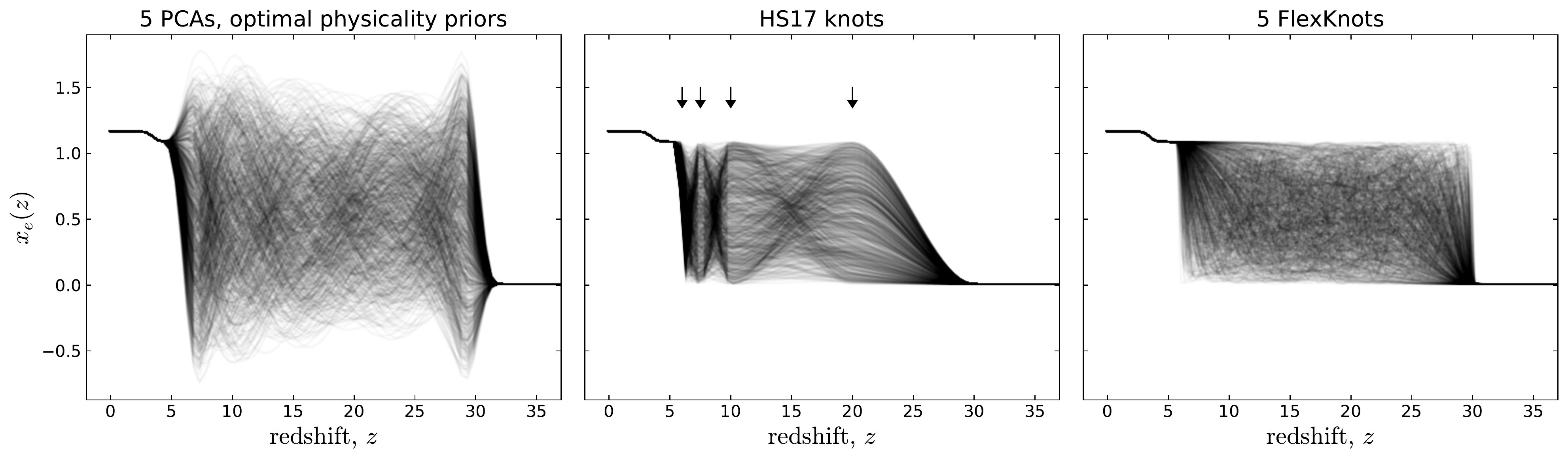}
\caption{One thousand reionization histories sampled from their prior distribution given the different models we consider. In the PCA case, the prior over PCA modes is taken to be uniform within the physicality region; in the two knot cases, the prior is uniform on the position and/or amplitude of each knot (thus none of these priors correspond to a flat $\tau$ priors). The left panel shows the extent to which the PCA procedure allows unphysical models, even when using the optimal physicality priors. The middle panel shows that the HS17 prior creates odd features and tighter peaked priors in the regions between the knot locations (which are indicated with arrows). The right panel is the prior for the \FlexKnot model and represents our best solution for creating a reasonable and ``homogeneous'' prior.}
\label{fig:reio_hists_prior}
\end{figure*}

As discussed in the previous section and elsewhere \citep[e.g.,][]{mortonson2008}, one draw-back of the PCA model is the necessary inclusion of unphysical parameter space in the prior. In some sense this is because the PCA procedure rotates the parameters from a space in which the prior is easy to describe to one in which the likelihood is easy to describe. If we are in a regime where the prior is completely uninformative, this is a very beneficial rotation; here, however, this is not quite the case. 

One solution is thus to not perform the PCA rotation at all, and keep the values of ionization fraction at some given redshifts (the $x_i$) as the parameters. Then we can always fully and sufficiently enforce physicality by requiring that $0\,{<}\,x_i\,{<}x_{\rm e}^{\rm max}$ for all $z$. Variations on this avenue have been explored in e.g. \cite{colombo2009}, \cite{lewis2006}, and \cite{hazra2017}. In this section we consider for definiteness the latter model, which we denote the HS17 model.

The HS17 model takes uniform priors on the values of $x_{\rm e}(z)$ at $z$ = 6, 7.5, 10, and $20$, with end-points at $x_{\rm e}(5.5)\,{=}\,x_{\rm e}^{\rm max}$ and $x_{\rm e}(30)\,{=}\,0$ fixed, and interpolates everywhere in between these knots using piecewise cubic Hermite interpolating polynomials (PCHIP). Although the choice is ad-hoc and made somewhat a posteriori (which makes judging the statistical evidence for this model difficult), it does likely capture much of the observable features of the reionization history. A larger problem, however, is that it creates an undesired prior distribution, shown in the middle panel of Fig.~\ref{fig:reio_hists_prior}. While the prior on $x_{\rm e}(z)$ at the $z$-values of the knots is uniform, it becomes triangular at the mid-point between knots, leading to the odd patterns seen in this figure. With a simple modification of this model, which we give in the next section, we are able to remove this oddly patterned prior.

\subsection{FlexKnots}

To remedy the non-idealities of both the PCA and HS17 models discussed thus-far, we develop the following model and analysis procedure, inspired by the primordial power spectrum reconstruction in the \Planck analysis \citep{vazquez2012,planck2014-a24}. 

The model, which we call the \FlexKnot model, has knots which can move left and right in addition to up and down. Our parameters are thus a set of $x_i$ and $z_i$, with uniform priors across the ranges
\begin{gather}
    6<z_i<30,\\
    0<x_i<x_{\rm e}^{\rm max}.
\end{gather}
Additionally, given any set of $z_i$, we compute the reionization history by first sorting them before interpolating between the knots \citep[see also Appendix A2 of][]{handley2015}. We perform interpolation using the PCHIP scheme, similarly as to what is done in HS17.

Samples from this prior with 5 knots are shown in the right-most panel of Fig.~\ref{fig:reio_hists_prior}. We see that the prior distribution is much more uniform than the HS17 prior, with no clustering around any particular redshift or ionization fraction value. This is the effect of the left/right freedom of knots, coupled with the sorting procedure.

A final question for the \FlexKnot model remains, mainly how many knots to allow? We follow \cite{planck2014-a24} in marginalizing over the number of knots by computing the Bayesian evidence for each of $n$ knots,
\begin{align}
    \mathcal{Z}_n = \int d\theta \, \mathcal{L}_n(\theta\,|\,d) \mathcal{P}_n(\theta),
\end{align}
where $\mathcal{L}_n$ and $\mathcal{P}_n$ are the likelihood and prior given $n$, and computing the posterior distribution marginalized over $n$,
\begin{align}
    \mathcal{P}(y) = \sum_{n=1}^N \frac{\pi_n \mathcal{P}_n(y)}{\mathcal{Z}_n}.
\end{align}
Here, $y$ can represent the ionization fraction history, $\tau$, or any other derived quantity of interest.  The $\pi_n$ are the prior weights which we give to a model with $n$ knots; we set this equal to one, i.e, we assume that every number of knots is equally probable. The evidence computation is performed in practice with \textsc{PolyChord} \citep{handley2015a,handley2015}.

Given the necessity of choosing the $\pi_n$, one should ask: have we really gained anything by this procedure, or have we just swapped the requirement of making one choice of prior for another? To answer this, note that although some choice still remains, we have reduced the problem of picking a functional prior over all possible histories to one of simply picking a prior over which integer number of knots to take. This is a massive reduction in how arbitrary a choice we are required to make, and represents the key strength of this procedure. Additionally, as we will show in the next section, the data themselves largely disfavor the need for anything beyond one or two knots anyways, thus we are not particularly sensitive to the $\pi_n$, as long as a reasonable choice is made.

\section{Implicit prior on $\tau$}
\label{sec:implicit_tau}

\begin{figure}
\includegraphics[width=\columnwidth]{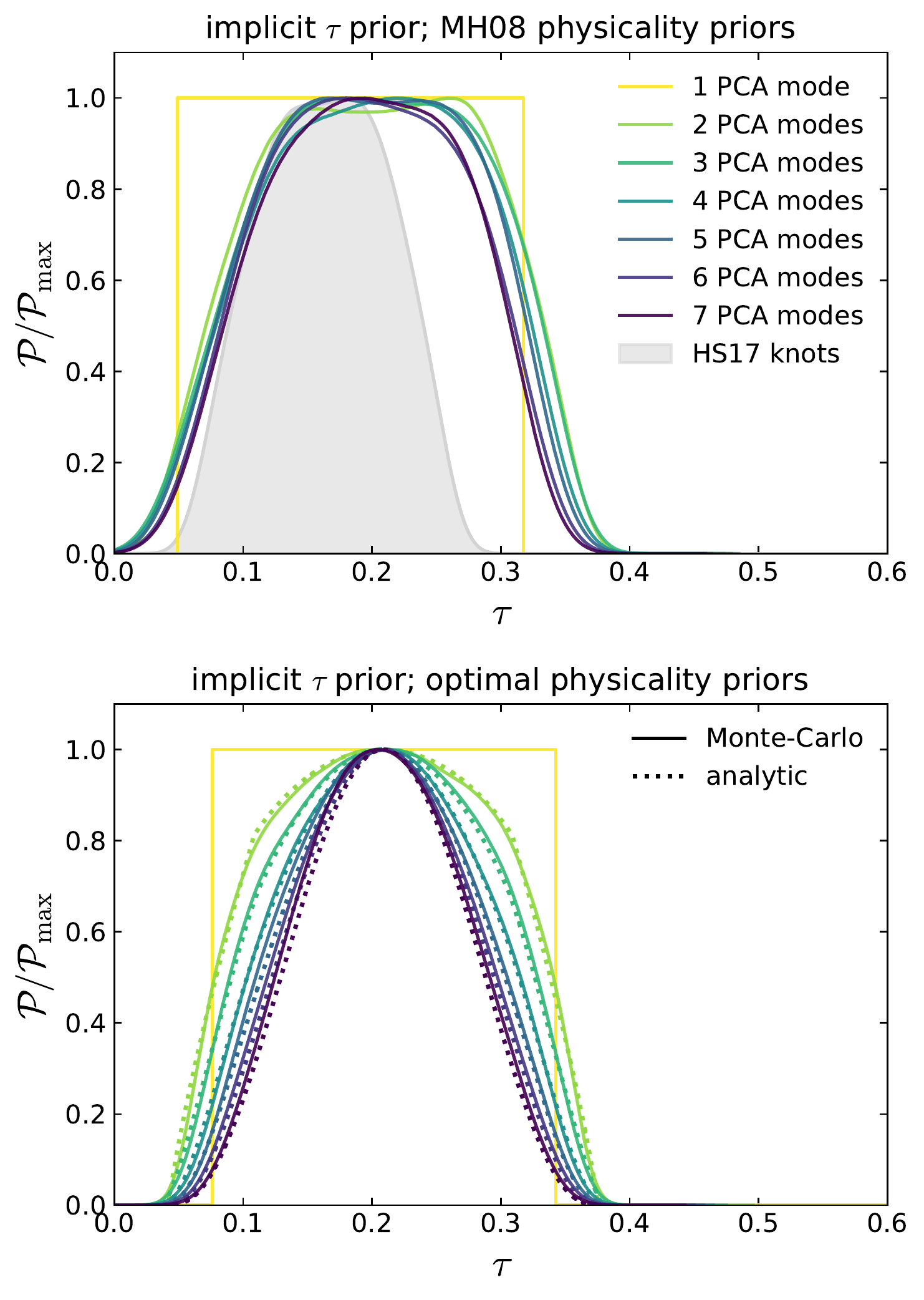}

\caption{The solid colored lines show the implicit prior on $\tau$ from the PCA model when taking a flat prior on the mode amplitudes, computed via Monte Carlo. The top panel assumes the MH08 physicality region, whereas the bottom panel assumes our optimal physicality region. In the case of the bottom panel, analytic calculation of the prior is possible via Eq.~\eqref{eq:tau_prior_analytic} and shown as dotted lines. The top panel also shows the implicit prior for the HS17 model.}

\label{fig:implicit_tau_prior}

\end{figure}

\subsection{{Calculating the prior}}
\label{sec:calc_tau}

Although the details of how reionization proceeds are of significant physical interest, it is mostly the total overall integrated optical depth to reionization, $\tau$, that plays an important role in terms of CMB constraints. This is because $\tau$ modulates the amplitude of temperature and polarization power spectra at multipoles greater that a few tens by $e^{-2\tau}$, and is thus a source of degeneracy for the scalar amplitude, $A_{\rm s}$, which does the same. The parameter $A_{\rm s}$ itself is then degenerate with a number of other physical quantities of interest via various means \citep{planck2016-LI}. It is thus important to examine what these different methods of reconstruction have to say in terms of $\tau$. 

In particular, what prior over $\tau$ do the different approaches assume? Take, for example, the PCA model. Because the modes contribute linearly to $x_{\rm e}(z)$ and because $\tau$ is in turn a linear functional of $x_{\rm e}(z)$, each mode contributes linearly to the total optical depth,
\begin{align}
\label{eq:taumodes}
\tau\big(\{m_i\}\big) = \tau_{\rm fid} + \sum_i \left(\frac{d\tau}{dm_i} \right) m_i,
\end{align}
where $\tau_{\rm fid}$ is the optical depth corresponding to $x_{\rm e}^{\rm fid}$ and
\begin{align}
\frac{d\tau}{dm_i} = \sigma_T \int dz \,\frac{n_{\rm H}(z)(1+z)^2}{H(z)} S_i(z).
\end{align}

In all analyses to-date, the prior taken on the $m_i$ has been uniform inside of the physicality region. We can compute the $\tau$ prior induced by this choice by propagating the $m_i$ prior to $\tau$ via Eq.~\eqref{eq:taumodes}. If the $m_i$ are uncorrelated in the prior (which is a decent approximation if using the MH08 physicality priors where the hypersphere prior plays a sub-dominant role) the induced $\tau$ prior will be a convolution of a number of ``top-hat'' functions, which should tends towards some smooth distribution peaking near the $\tau$ corresponding to $m_i\,{=}\,(m_i^+ + m_i^-)/2$. The exact solution (including both hyperbox and hypersphere) can be obtained numerically by Monte Carlo sampling from the $m_i$ prior and computing $\tau$ for each sample. This is shown in the top panel of Fig.~\ref{fig:implicit_tau_prior}. Indeed, we find that the prior is centered on $\tau\approx0.2$ and has width around $\sigma(\tau)\approx0.07$, shrinking and smoothing out as more modes are added. 

It is important to note the existence of this prior when comparing constraints on $\tau$ from two different models, for example the TANH model (which takes a flat $\tau$ prior) and the PCA model. In such comparisons, we change not only the model but also implicitly change the prior. Given that the prior we see in Fig.~\ref{fig:implicit_tau_prior} tends towards quite high values of $\tau$, it should not be entirely surprising if we find a higher $\tau$ posterior in the PCA case. In Sec.~\ref{sec:bflike} we will quantify the impact this has on the \cite{heinrich2017} analysis. 

What about the induced prior in the case of the optimal physicality prior instead of the MH08 one? Here, the extra symmetries of the problem allow us to derive the following useful analytic solution. First, consider the simplified case of computing the $\tau$ prior induced by only the hypersphere prior. This is given by
\begin{align} \label{eq:Ptau_int}
    \mathcal{P}(\tau) = \int_{\mathbb{S}}\; \displaystyle \prod_{i=1}^{N} dm_i \;\delta\left[\tau - \tau\big(\{m_i\}\big)\right],
\end{align}
where the region $\mathbb{S}$ is an $N$-dimensional hypersphere with radius $r\equiv x_{\rm e}^{\rm max}/2$ and centered on the origin, and $\delta$ is the Dirac $\delta$-function. In geometric terms, Eq.~\eqref{eq:Ptau_int} calculates the volume of the intersection of the hyperplane defined by $\tau - \tau(\{m_i\})=0$ with the hypersphere $\mathbb{S}$. This intersection is itself an $N-1$ dimensional hypersphere. Its radius, $\rho$, is smaller than $r$ due the hyperplane having sliced through $\mathbb{S}$ at a position displaced from its center. The displacement distance is controlled by $\tau$ and given by $D = |\tau-\tau_{\rm fid}|/|\mathbf{g}|$, where we have defined a vector normal to the hyperplane, $(\mathbf{g})_i \equiv d\tau/dm_i$. This allows us to compute the smaller radius, $\rho^2 = r^2 - D^2$, and, using the formula for the volume of a hypersphere, we arrive at the final answer,
\begin{align} \label{eq:Ptau_sphere}
    \mathcal{P}(\tau) \propto \frac{\pi^{\frac{N-1}{2}}}{\Gamma(\frac{N}{2}-1)}  \rho^{N-1} \propto \left[1-\frac{(\tau-\tau_{\rm fid})^2}{r^2|\mathbf{g}|^2}\right]^{\frac{N-1}{2}}.
\end{align}
Note that, as written, neither of these forms are a properly normalized probability distribution (however, this is not a requirement for our purposes, and the proper normalization is straightforward to compute if desired). 

We now need to include the fact that the hyperbox prior excludes some of the hypersphere prior, as represented in two dimensions in the right panel of Fig.~\ref{fig:hyperbox}. In general, this breaks the symmetries which made Eq.~\eqref{eq:Ptau_sphere} so simple. We find, however, that because $d \tau/d m_1$ is much larger than any of the other derivatives, it is predominantly only the exclusion in the $m_1$ direction which needs to be accounted for. This leaves enough symmetry that the resulting solution is simple enough to be useful.

The above calculation needs to be amended in the following way. Define $\mathbb{S}^\prime$ to be intersection of the hypersphere with the region defined by $m^{-}_1\,{<}\,m_1\,{<}\,m^{+}_1$; i.e., $\mathbb{S}^\prime$  is a hypersphere with two opposite hyper-spherical caps removed. As $\tau$ is increased or decreased from $\tau_{\rm fid}$, changing where the hyperplane intersects with $\mathbb{S}^\prime$, the intersection region initially does not also intersect with the caps, leaving the solution unchanged from before. Once it does intersect with the caps, the intersection region is now no longer an $N-1$ dimensional hypersphere, but rather an $N-1$ dimensional hypersphere with a single hyper-spherical cap removed. Thus, the induced prior will be instead
\begin{align} \label{eq:tau_prior_analytic}
    \mathcal{P}(\tau) \propto \frac{\pi^{\frac{N-1}{2}}}{\Gamma(\frac{N}{2}-1)}  \rho^{N-1}(1 - V),
\end{align}
where V is the volume of the removed hyper-spherical cap relative to the volume of the entire hypersphere. This fraction can be written in terms of the regularized incomplete beta function, $I_x(a,b)$ \citep{li2011}, and is given by
\begin{align}
   V = \begin{cases} 
      0 & h<0 \\
      \frac{1}{2} \, I_{(2\rho h-h^2)/\rho^2}\left(\frac{N}{2},\frac{1}{2}\right) & 0<h<\rho \\
      1-\frac{1}{2} \, I_{(2\rho h-h^2)/\rho^2}\left(\frac{N}{2},\frac{1}{2}\right) & h>\rho,
   \end{cases}
\end{align}
where $h$ is the height of the cap, $h=\rho-m^{(+)} \csc \theta + D \cot \theta$, and $\theta$ is the angle between the hyperplane and the $\hat m_1$ direction, $\cos \theta = \mathbf{g}_1/|\mathbf{g}|$.

The bottom panel of Fig.~\ref{fig:implicit_tau_prior} shows this analytic result for several values of $N$, alongside the Monte Carlo calculation performed similarly as before. The ``kink'' near $\tau\,{\sim}\,0.1$ (visible particularly in the $N\,{=}\,2$ curve, although present in all of them) corresponds to when the correction, $V$, turns on; without this, we would not recover the correct shape in the tails of the distribution. With it, however, we find very good agreement between the Monte Carlo simulations and the analytic result, particularly deep in the low-$\tau$ tail which is most important since this is the region picked out by the data. This analytic result will be useful in the next section where we discuss flattening the prior. Note also that the prior volume for $\tau$ shrinks noticeably as compared to the MH08 physicality region; we will show in Sec.~\ref{sec:bflike} that this smaller and more optimal prior volume is enough to affect constraints from \Planck data.

Similarly as to the PCA model, both the HS17 knot model and our \FlexKnot model give a non-flat $\tau$ prior if the prior on the knot amplitudes and/or positions is taken as flat. The prior for the HS17 model is shown in the top panel of Fig.~\ref{fig:implicit_tau_prior}. The \FlexKnot prior is not shown, but is qualitatively similar to the others.

\subsection{Flattening the prior}
\label{sec:flatten_tau}

Having ascertained that all of the generic reconstruction methods we have discussed implicitly place a non-flat prior on $\tau$, we now discuss how to modify these analyses so that any desired $\tau$ prior can be used, in particular a uniform one. This will be useful in performing more consistent comparisons between models by using the same prior on $\tau$ for all models.

Intuitively, if we have an MCMC chain for an analysis that was performed with some particular $\tau$ prior, we can importance sample the chain to give additional weight to samples with a low prior probability for $\tau$, in effect counter-balancing the original prior and forcing it to be flat. More explicitly, we can create a posterior where we have assumed a flat $\tau$ prior by computing
\begin{align} \label{eq:flatten_tau}
    \mathcal{P}^{{\rm flat}\text{-}\tau}(\theta\,|\,d) = \frac{\mathcal{P}^{\rm orig}(\theta\,|\,d)}{\mathcal{P}^{\rm orig}(\tau(\theta))} \propto \mathcal{L}(d\,|\,\theta) \underbrace{\left[\frac{\mathcal{P}^{\rm orig}(\theta)}{\mathcal{P}^{\rm orig}(\tau(\theta))}\right]}_{\mathcal{P}^{{\rm flat}\text{-}\tau}(\theta)},
\end{align}
where the posterior $\mathcal{P}^{\rm orig}(\theta\,|\,d)$ given data $d$ is the original posterior that did not assume a flat $\tau$ prior, $\mathcal{P}^{\rm orig}(\theta)$ and $\mathcal{P}^{\rm orig}(\tau(\theta))$ are the original priors in terms of all parameters $\theta$ and the induced $\tau$ prior, respectively, and $\mathcal{L}(d\,|\,\theta)$ is the likelihood of the data given parameters. In the case of the PCA model with optimal physicality priors, $\mathcal{P}^{\rm orig}(\tau(\theta))$ is conveniently calculable analytically and given in Eq.~\eqref{eq:tau_prior_analytic}. For other models, it can be obtained via Monte Carlo simulations and a smooth function can be fit \citep[e.g., here we use kernel density estimates with tools from][]{antonylewis2016}.

We should note that this procedure is not unique, and many $\mathcal{P}^{{\rm flat}\text{-}\tau}(\theta)$ exist corresponding to a flat prior on $\tau$. However, it can be rigorously shown that the choice we have made in Eq.~\eqref{eq:flatten_tau} maximizes the possible entropy in $\mathcal{P}^{{\rm flat}\text{-}\tau}(\theta)$, subject to the condition that the resulting $\tau$ prior is flat \citep[][we also give a simplified but less general proof of this in Appendix ~\ref{app:flattau}]{handley2018}. This is to say that Eq.~\eqref{eq:flatten_tau} is very well motivated because it is the choice that takes the original prior and modifies it in such a way that the induced $\tau$ prior is flat while introducing the minimal amount of new information. We thus make use of this procedure in the following section.

\section{Hints of early reionization in Planck 2015 data?}
\label{sec:bflike}

\begin{figure}  
\includegraphics[width=\columnwidth]{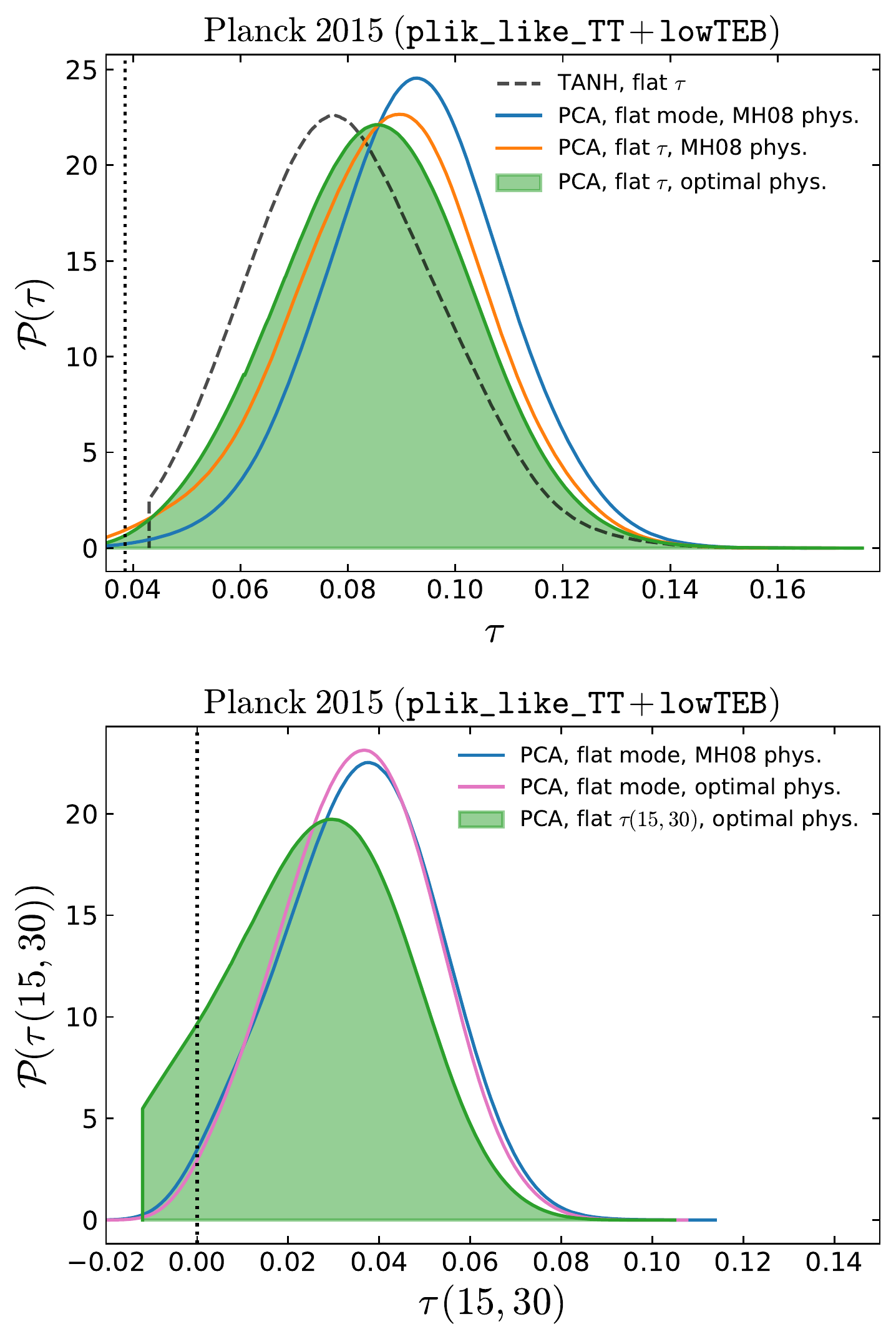}

\caption{Constraints on the total optical depth (top panel), and on the optical depth between $z\,{=}\,15$ and $z\,{=}\,30$ (bottom panel), when using 2015 PlanckTT+lowTEB data. Various reionization models and priors are used (labeled in each plot), in addition to marginalization over other $\Lambda$CDM parameters. The top panel shows that the apparent preference for higher $\tau$ when using the PCA model as reported in \cite{heinrich2017} was significantly impacted by the choice of prior. The bottom panel shows that evidence for early reionization was also partly impacted, although some preference remains. We consider the shaded green contour, which corresponds to a flat $\tau$ prior and optimal physicality region, as the most well-motivated constraint from this dataset.}

\label{fig:data_plikTT_bflike}
\end{figure}

In this section, we consider the impact of the priors discussed thus far on claims in \cite{heinrich2017} and \cite{heinrich2018} for evidence of early reionization in the \Planck 2015 data. 

\cite{heinrich2017} show that the value of $\tau$ inferred from the \Planck 2015 data is almost $1\,\sigma$ higher under the PCA model than with the TANH model. We reproduce\footnote{One small difference is that \cite{heinrich2017} use the full TT, TE, and EE data, whereas here we use only the ``robust'' TT data. We have checked that this makes a very small difference to the conclusions in this section, as expected since the TT data is much more constraining.} this result in the top panel of Fig.~\ref{fig:data_plikTT_bflike}, where the black and blue lines there correspond to the black and blue lines in Fig.~5 of \cite{heinrich2017}. When we use a flat prior on the modes and the MH08 physicality region as they have, we find very good consistency with their results. We expect that some of the shift to higher $\tau$ is due to the high $\tau$ prior implicit in the flat mode prior; to quantify the exact effect we flatten the $\tau$ prior using the procedure described in the previous section and show the result in orange. Indeed, the shift to higher $\tau$ is partly reduced. We also switch to the optimal physicality region, and show the final result in shaded green. This further reduces the inferred value of $\tau$, which now agrees much better with the TANH result. We explicitly find
\begin{align}
    \tau = 0.080\pm0.017\;\;{\scriptstyle(\rm TANH,\;flat\,\tau;\;P15)}
\end{align}
and 
\begin{align}
    \tau = 0.085\pm0.018\;\;{\scriptstyle(\rm PCA,\;flat\,\tau,\;optimal\,phys.;\;P15)}
\end{align}
which is a shift of only $0.3\,\sigma$. We stress that the difference between this and the nearly $1\,\sigma$ shift found by \cite{heinrich2017} is due purely to a different choice of priors over reionization histories inherent in the PCA procedure. 

Although highlighting the shift in $\tau$, \cite{heinrich2017} do not make the claim that one should regard the value of $\tau$ inferred from the PCA procedure as model independent; the results of the previous paragraph should make that point very clear. However, a strong claim is made, further argued in \cite{heinrich2018} and \cite{obied2018}, that the PCA procedure provides reionization-model-independent proof of {\it early} reionization because the value of $\tau(15,30)$ is greater than zero at $2\,\sigma$. 

The derived parameter $\tau(15,30)$ is not fundamentally different from the derived parameter $\tau$, and thus its posterior distribution will qualitatively depend on our choice of priors in the same way. In the bottom panel of Fig.~\ref{fig:data_plikTT_bflike} we perform a similar set of tests for $\tau(15,30)$ as we just have for $\tau$. The blue line is the PCA result with flat mode prior and the MH08 physicality region. We find it is higher than zero by $2.1\sigma$, again in good agreement with the \cite{heinrich2017} result. We switch to the optimal physicality region in purple and additionally to a flat prior on $\tau(15,30)$ in shaded green, yielding,
\begin{gather}
    \label{eq:tauearly_bflike}
    \tau(15,30) = 0.027\pm0.019\\
    {\scriptstyle(\rm PCA,\;flat\,\tau(15,30),\;optimal\,phys.;\;P15)}\nonumber
\end{gather}
This represents a decrease in the evidence for non-zero $\tau(15,30)$ from $2.1\,\sigma$ to $1.4\,\sigma$. Thus, similarly as before for $\tau$, we do not find robustness to choice of priors in this detection of early reionization. Evidently, part of the evidence for early reionization found by \cite{heinrich2017} is actually due to the choice of prior rather than being driven by the data. 

\cite{heinrich2018} make the contrary argument, claiming that the PCA evidence for early reionization is robust to the choice of prior. Here, we point out some problems with the arguments therein. First, they show that increasing the $z_{\rm max}$ of the reconstruction does not reduce preference for early reionization, arguing that this shows priors are uninformative. In fact, their results actually highlight the opposite. In particular, increasing the $z_{\rm max}$ of the reconstruction increases the mean value of the prior on $\tau(z,z_{\rm max})$ for all $z$. As a result, one would expect the mean value of the posterior of this quantity to increase with higher $z_{\rm max}$, which is exactly what is seen in all cases in Table I of \cite{heinrich2018}. Furthermore, they perform a likelihood ratio test (which is, by definition, independent of priors), finding that the $\chi^2$ is decreased by 5--6 with a 5-parameter PCA model as compared to the 1-parameter TANH model. Without a quantitative argument as to the effective degrees of freedom of the PCA model that are constrained by the data (which is not made), it is impossible to judge the significance of the decrease in $\chi^2$. Conversely, we consider the very direct tests performed here of computing and flattening the prior to be more conclusive as to the impact of these priors on \Planck 2015 data. 

Of course, some small hints of non-zero $\tau(15,30)$ remain even in the case of the constraints shown in Eq.~\eqref{eq:tauearly_bflike}. However, we will show in the next section that these hints are consistent with arising from a noise fluctuation, since they disappear almost entirely with the addition of the newest \Planck large-scale polarization data.

\section{Reionization from Planck intermediate data}
\label{sec:simlow}

\begin{figure}  
\includegraphics[width=\columnwidth]{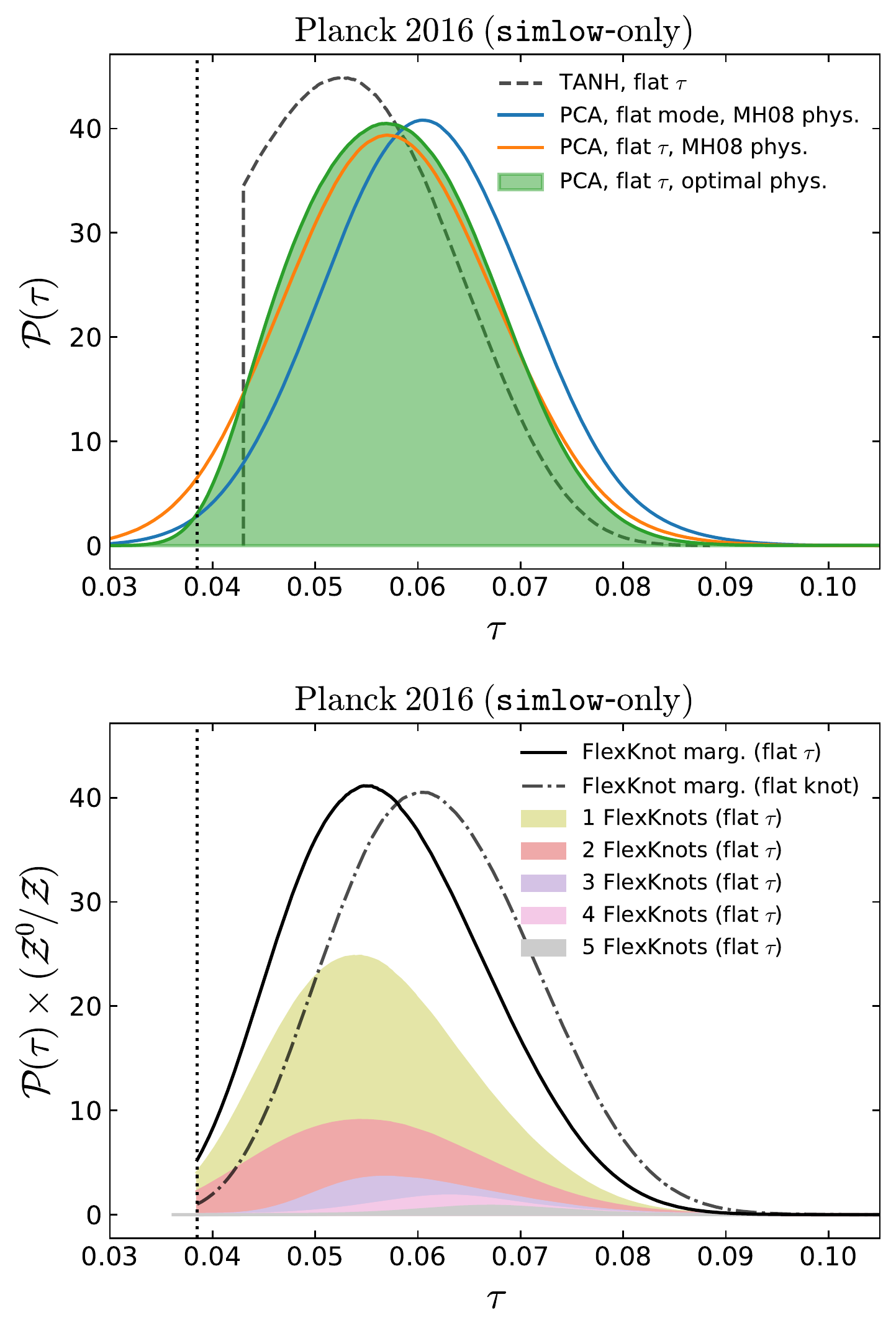}

\caption{Constraints on the total optical depth when using the \Planck intermediate \texttt{simlow} likelihood alone. Various reionization models and priors are used (labeled in each plot), with other $\Lambda$CDM parameters fixed to their best-fit values from PlanckTT data. In the bottom panel, the height of each \FlexKnot model is proportional to its Bayesian evidence, with an overall scaling so that their sum (the black curve) gives a properly normalized probability distribution. This curve corresponds to marginalization over the number of \FlexKnot to use.}

\label{fig:data_simlow}
\end{figure}

The \Planck polarization likelihood used in the previous section and by \cite{heinrich2017} and \cite{heinrich2018} is the most recent public \Planck likelihood available and is based on polarization maps from \Planck-LFI data \citep{planck2014-a03}. Reionization constraints from lower-noise HFI data were presented in an intermediate release \citep{planck2014-a10}. Additionally, \cite{planck2014-a25} also explored various parametric reionization models extending beyond just the TANH case. In general, good consistency with a near-instantaneous transition was found, and the optical depth constraints tightened and shifted to lower central values of $\tau\,{\approx}\,0.055$--$0.060$, with $\sigma(\tau)\,{\sim}\,0.01$. Although no evidence for early reionization was found, no completely generic reconstruction was performed, and the parametric models considered did not fully accommodate an early component should one have been present \citep{heinrich2017}. We thus give here results from a fully generic reconstruction of the reionization history using the \texttt{simlow} likelihood.

So as to most clearly judge the impact of the large-scale polarization data, we compute constaints using {\it only} \texttt{simlow}. To do so, we fix non-reionization cosmological parameters to their best-fit values from the high-$\ell$ TT data, in particular holding the quantity $10^9 A_{\rm s} e^{-2\tau}$ fixed (which better approximates the impact of the TT data as compared to holding just $A_{\rm s}$ fixed). The black dashed contour in the top panel of Fig.~\ref{fig:data_simlow} shows constraints on $\tau$ from \texttt{simlow} assuming the TANH model, as also presented in \cite{planck2014-a10} and \cite{planck2014-a25}. Note that a significant part of the posterior density is cut off by Gunn-Peterson bound requiring full reionization by $z\,{=}\,6$, which for the TANH model translates to $\tau\,{>}\,0.0430$. Due to the finite width of reionization assumed in the TANH case ($\Delta z\,{=}\,0.5$), this is slightly higher than the absolute minimum possible for instantaneous reionization at $z\,{=}\,6$, which is $\tau\,{>}\,0.0385$. The remaining lines show results using the PCA model and various choices of prior. As before, when flattening the $\tau$ prior we see a downward shift in $\tau$. Switching to the optimal physicality priors now mostly has the effect of removing some weight at $\tau\,{<}\,0.0385$, which before was unphysically allowed by the MH08 region. Switching to the optimal physicality region remedies some of this effect and gives
\begin{align}
    \tau = 0.057\pm0.009\;\;{\scriptstyle(\rm PCA,\;flat\,\tau,\;optimal\,phys.;\;\mathtt{simlow})}
\end{align}
This is in good agreement with the TANH result, as well as those from the parametric models of \cite{planck2014-a25}.

In the bottom panel of Fig.~\ref{fig:data_simlow}, we show constraints from the \FlexKnot model. In all cases except for the dot-dashed line, we use a flat prior on $\tau$. The posteriors on $\tau$ with varying numbers of knots are shown as the shaded contours. Each posterior has been normalized to unity then multiplied by $\mathcal{Z}^{0}/\mathcal{Z}_i$, where $\mathcal{Z}_i$ is the evidence for that number of knots and $1/\mathcal{Z}^{0}=\sum_i 1/\mathcal{Z}_i$. 
This means that the height of each curve is proportional to the evidence of that model, and makes it easy to see that the 1-knot model has most evidence, and the evidence decreases for all subsequent models. Summing up these curves produces the posterior on $\tau$ marginalized over the number of knots, which is shown in black, and is properly normalized to unity by the choice of $\mathcal{Z}^{0}$. In this case we find
\begin{align}
    \label{eq:tau_simlow_flattau}
    \tau = 0.058\pm 0.009\;\;{\scriptstyle (\rm FlexKnot,\;flat\,\tau;\;\mathtt{simlow})}
\end{align}
This as well is in very good agreement with the results above, demonstrating the lack of evidence in the \texttt{simlow} data for anything other than a single late and near-instantaneous transition to a fully ionized universe. 

\begin{figure}
\includegraphics[width=\columnwidth]{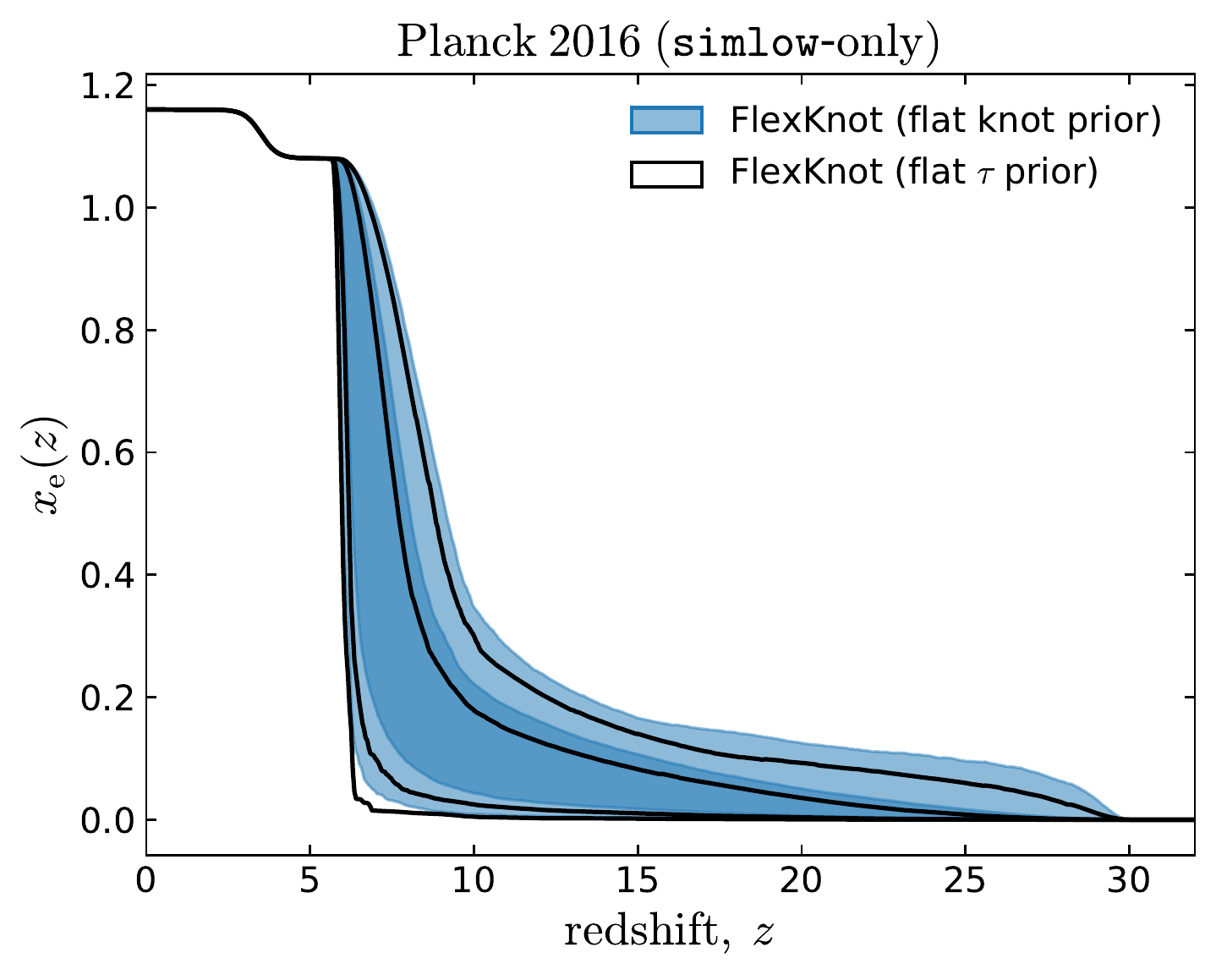}

\caption{Constraints on the ionization fraction as a function of redshift from \texttt{simlow}-only data, using either a flat prior on the knot positions or a flat prior on $\tau$. The blue and black contours represent the middle 68\% and 95\% quantile of the posterior at each redshift.}

\label{fig:reio_hists_simlow}
\end{figure}

Even more directly, we show in Fig.~\ref{fig:reio_hists_simlow} the posterior constraints on the history itself using the \FlexKnot model, marginalized over the number of knots. Here we see the tight restriction on any amount of non-zero ionization fraction at high redshifts, independent of exactly which $\tau$ prior we use. Indeed, the posterior on the $z\,{>}\,15$ contribution to the optical depth is bounded to be
\begin{gather}
    \label{eq:tauearly_simlow}
    \tau(15,30) < 0.015 \;(95\%) \\
    {\scriptstyle (\rm FlexKnot,\;flat\,\tau(15,30);\;\mathtt{simlow})}\nonumber
\end{gather}
Although we consider the \FlexKnot result more robust than the PCA one, we also compute the PCA bound on this quantity to stress that the disappearance of the hints of early reionization in the intermediate \Planck data is more related to the data rather than the model used. Even in the PCA case, we find
\begin{gather}
    \tau(15,30) < 0.028 \;(95\%) \\
    {\scriptstyle (\rm PCA,\;flat\,\tau(15,30),\;optimal\,phys.;\;\mathtt{simlow})}\nonumber
\end{gather}
which is weaker but still largely rules out the central value of even the prior-adjusted results given by \Planck 2015 data (Eq.~\ref{eq:tauearly_bflike}).

Of course, the \FlexKnot model is not immune to choice of $\tau$ prior. For comparison, we show as the dot-dashed line in Fig.~\ref{fig:data_simlow} the \FlexKnot result, marginalized over knots, but with a flat prior on the knot positions and amplitudes instead of a flat prior on $\tau$. We find
\begin{align}
    \tau = 0.062\pm 0.009\;\;{\scriptstyle (\rm FlexKnot,\;flat\,knot;\;\mathtt{simlow})}
\end{align}
which is $0.4\,\sigma$ higher than the result with a flat $\tau$ prior (Eq.~\ref{eq:tau_simlow_flattau}). We regard this difference as quantifying an unavoidable systematic uncertainty in $\tau$ due to imperfect knowledge of the model parameter priors which is inherent in fully generic reconstructions of the reionization history. In the next section, we consider the impact of this uncertainty on other cosmological parameters that are correlated with $\tau$, both from current and future data.

\section{Impact on Future Data}
\label{sec:future}

We now turn to forecasting the impact of reionization uncertainty on future data. There are two question we would like to answer. The first is whether the increased errors bars on $\tau$ when performing a generic reconstruction cause any significant degradation in constraints on other parameters of interest. In the case of Fig.~\ref{fig:data_simlow} we see around a 25\% larger error bar in the \FlexKnot case as compared to the TANH case\footnote{We compute the TANH error bar here by taking the standard deviation the TANH posterior. Despite that this is clearly non-Gaussian as it is cut off by the Gunn-Peterson prior, it gives us a rough estimate to use in combination with Fisher forecasts.}. We will check the impact of this degradation, as well as forecasting whether this can be improved upon with better data. Secondly, we will take the shift in the $\tau$ posterior between a flat knot prior and a flat $\tau$ prior and propagate it into shifts on other parameters of interest given future data, to check the extent to which these future constraints are prior-dependent. In our discussion, we focus on the sum of neutrino masses, $\Sigma m_\nu$, as it is the cosmological parameter expected to be most impacted by the details of reionization \citep{allison2015}.

The first dataset we consider is a combination of CMB-S4 with existing \Planck data (including the intermediate \texttt{simlow} likelihood). We perform a Fisher forecast using the standard procedure \citep{dodelson2003}, using a fiducial \Planck 2015-like cosmology, with $\tau\,{=}\,0.055$ and $\Sigma m_\nu$ at the minimum value of $60\,\rm meV$. For CMB-S4, we assume a temperature noise level $\Delta_T =\Delta_{P}/\sqrt{2} = 1\,\rm \mu K$-arcmin, a fraction of covered $f_{\rm sky} = 0.5$, and a beam size of $\theta_{\rm FWHM} = 3\,\rm arcmin$. For the noise levels of the reconstructed lensing deflection power spectrum, we use the noise power spectrum calculated by \cite{pan2015}\footnote{Provided by the authors upon request.} based on an iterative quadratic EB estimator \citep{okamoto2003,smith2012}. We assume the largest scales measurable by CMB-S4 are $\ell_{\rm min}\,{=}\,50$. This is combined with the constraints on $\tau$ from \texttt{simlow} discussed in the previous section, which are treated as independent. In combining with the remaining \Planck 2015 data, to take into account correlations between \Planck and CMB-S4 due to sky coverage and multipole overlap, we compute a \Planck-like Fisher forecast rather than use the real \Planck constraints. The temperature and polarization noise levels assumed for \Planck are those given by \cite{planck2005-bluebook}, taking an optimal noise weighting of the 100, 143, and 217\,GHz, channels, and a sky coverage of $f_{\rm sky}\,{=}\,0.75$. These reproduce the uncertainties on parameters from the real \Planck data to within around 15\%, which should be good enough for our purposes.

\begin{figure}
\includegraphics[width=\columnwidth]{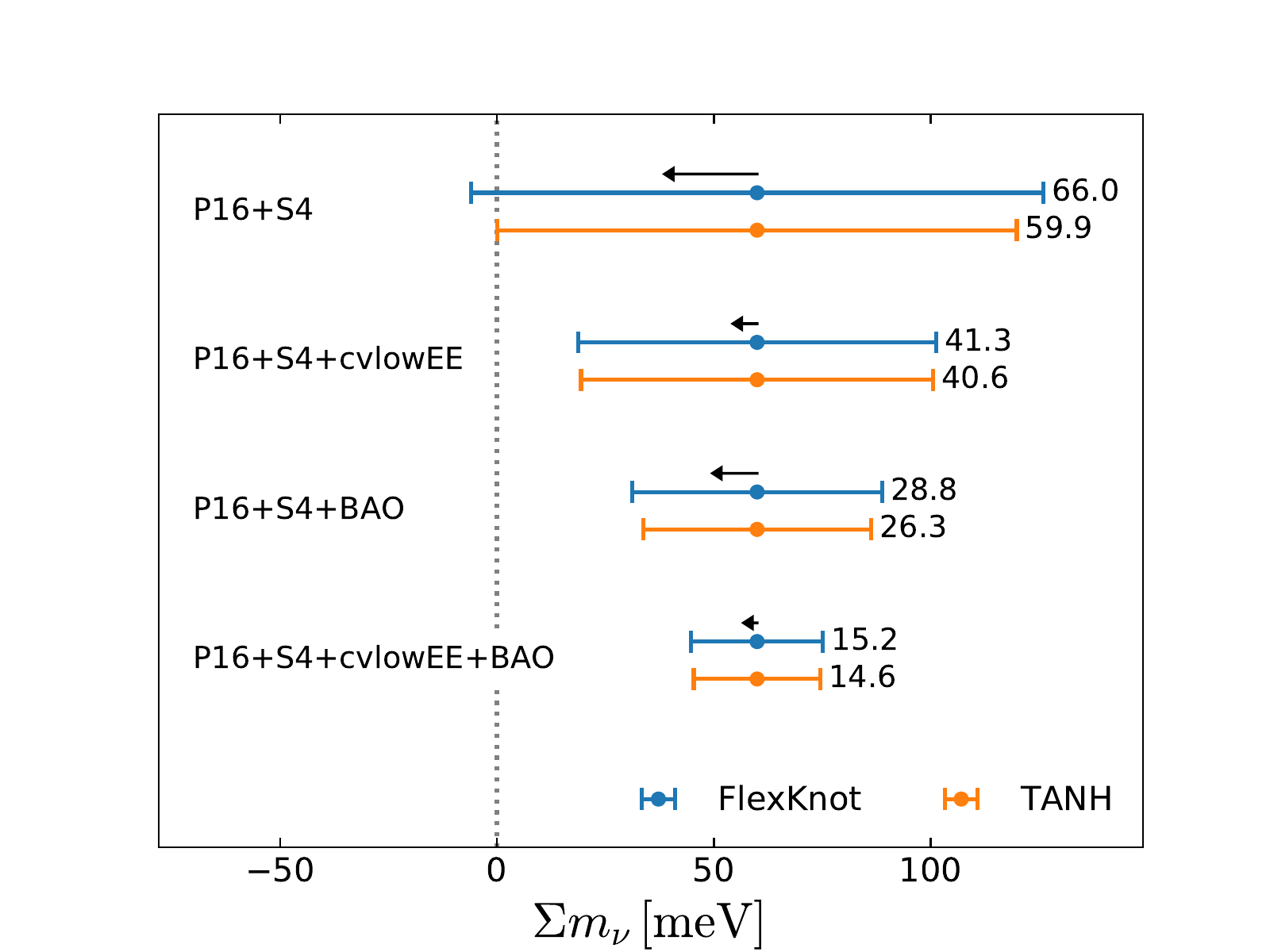}

\caption{Fisher forecasted $1\,\sigma$ error bars on the sum of neutrino masses, $\Sigma m_\nu$, assuming a fiducial value at the minimum allowed sum of $60\,\rm meV$. Different datasets are labeled on the left, with P16 corresponding to \Planck 2015 data combined with the intermediate \texttt{simlow} likelihood, S4 corresponding to CMB-S4, BAO to DESI-BAO, and \texttt{cvlowEE} to a cosmic-variance-limited full-sky EE measurement up to $\ell\,{=}\,30$. Orange error bars use the TANH reionization model, and blue errors bars are slightly degraded due to having marginalized over all possible reionization histories via the \FlexKnot model. The black arrows show the expected shift when considering a flat knot prior vs. a flat $\tau$ prior.}
\label{fig:mnu}
\end{figure}

For this first dataset, we forecast a $1\,\sigma$ error bar on $\Sigma m_\nu$ of 59.9\,meV assuming the TANH model, or 66.0\,meV with the 25\% looser $\tau$ prior coming from the \FlexKnot model. This fairly modest degradation is depicted by the error bars labeled P16+S4 in Fig.~\ref{fig:mnu}. What about the impact from the $0.4\,\sigma$ shift in $\tau$ depending on the prior assumed? The Fisher methodology allows us to propagate this into a shift in $\Sigma m_\nu$ via
\begin{align}
    \frac{\Delta m_\nu}{\sigma_{m_\nu}} = \rho \frac{\Delta \tau}{\sigma_\tau},
\end{align}
where $\rho$ is the correlation coefficient between $\tau$ and $\Sigma m_\nu$. For the P16+S4, we find $\rho\,{=}\,0.8$, thus the shift in $\Sigma m_\nu$ is $0.3\,\sigma$. This is depicted by the black arrow in Fig.~\ref{fig:mnu}. It is arbitrarily chosen to point to lower $\Sigma m_\nu$, which is the direction of shift we would expect when going from a flat knot prior to a flat $\tau$ prior. 

The existence of this shift is a main conclusion from this work, and has not been considered before. To the extent that both flat knot and flat $\tau$ priors are reasonable, this can be considered as an extra source of ``model uncertainty'' in the future CMB determination of neutrino mass. The magnitude of the effect is not disastrous, but not negligible either. We comment now on a few avenues to reduce its impact. 

The first comes from adding in other expected future constrains from measurements of the baryon acoustic oscillations (BAO) feature in the galaxy correlation function by DESI \citep{levi2013}. We use the Fisher matrix for DESI calculated by \cite{pan2015}, based on the sensitivities presented in \cite{font-ribera2014}. The addition of this dataset brings us significantly closer to a guaranteed detection of the neutrino mass. Here we find a $1\,\sigma$ error bar on $\Sigma m_\nu$ of 26.3\,meV or 28.8\,meV, again depending on the reionization model used\footnote{These predictions are slightly less optimistic than the ones presented in the main body of \cite{pan2015}, mostly because we use $\ell_{\rm min}\,{=}\,50$ for CMB-S4 as opposed to $\ell_{\rm min}\,{=}\,2$ used there. They are in better agreement with \cite{allison2015} which take a similar $\ell_{\rm min}$, but still very slightly wider, due mostly to a lower fiducial value of $\tau$.}. Although the error bars shrink, the addition of DESI actually increases the correlation between $\tau$ and $\Sigma m_\nu$ slightly to $\rho\,{=}\,0.9$. This arises because DESI is completely insensitive to the $\tau$-$A_{\rm s}$ degeneracy responsible for the correlation, but reduces other physical degeneracies  impacting the $\Sigma m_\nu$ determination, thus leaving the former degeneracy more prominent. In this case, we find a $0.35\,\sigma$ shift in $\Sigma m_\nu$ depending on the $\tau$ prior used. While this is a larger relative shift, on an absolute scale it is about half the shift as compared to without DESI. 

To reduce $\rho$ and thus the impact of reionization uncertainty, we need data which directly constrains $\tau$ and/or $A_{\rm s}$. This could be provided, for example, by more precise large-scale polarization data. There is room for some improvement as the \Planck data is not yet at the cosmic variance limit. We thus consider the limiting case of a full-sky cosmic-variance limited EE measurement across multipoles 2 to 30 (which we will label \texttt{cvlowEE}). This should be the upper bound of what could be achieved with next-generation satellite missions, e.g., LiteBIRD, PIXIE, COrE, or PICO \citep{matsumura2016,kogut2011,thecorecollaboration2011}. To obtain the most accurate forecasts, we run MCMC chains using the exact full-sky $C_\ell$ likelihood \cite[e.g.][]{hamimeche2008}. This is more accurate than Fisher forecasts, which do not easily deal with the sharp edges of the physicality priors. Using the TANH model, we forecast an error bar of 
\begin{align}
    \sigma_\tau = 0.0020\;\;{\scriptstyle (\rm TANH,\;flat\,\tau;\;\mathtt{cvlowEE})}
\end{align}
and with the \FlexKnot model marginalized over the number of knots, we obtain
\begin{align}
    \sigma_\tau = 0.0024\;\;{\scriptstyle (\rm FlexKnot,\;flat\,\tau;\;\mathtt{cvlowEE})}
\end{align}
This leads to a 15\% degradation of the error bar in $\Sigma m_\nu$, somewhat better than the 25\% degradation we found previously with \texttt{simlow}. Additionally, $\rho$ is reduced to 0.5, meaning the shift in $\Sigma m_\nu$ due to the choice of $\tau$ prior is now only $0.2\,\sigma$. Evidently, these improved large-scale polarization measurements would be quite valuable, not just for reducing the absolute uncertainty on $\tau$, but also for reducing our dependence on its prior. 

When we combine all datasets discussed thus far, including both \texttt{cvlowEE} and DESI-BAO, we find an expected constraint of $\sim15\,\rm meV$, with the $\tau$ prior able to shift things only by  $\sim3\,\rm meV$. This would be enough for a  $\sim4\,\sigma$ guaranteed detection of non-zero neutrino masses even if the true value is at the minimum.

\section{Discussion and Conclusions}
\label{sec:comments}

In this work we have examined the impact of priors on generic reconstruction of the reionization history. We have pointed out some problems with PCA procedure, mainly that 1 the priors that are usually taken over the mode amplitudes correspond implicitly to a somewhat informative prior on $\tau$ and 2) a sub-optimal set of physicality priors has been used to-date. Ultimately, the dependence of the final posterior on the details of these physicality priors argues that the PCA procedure is not well suited to this problem. This is not to say there are issues with PCA analyses in general, simply that in situations where hard priors exist and are informative (such as here), the PCA method is a sub-optimal tool. Indeed, we have shown the significant extent to which some of these priors have impacted the analysis of \cite{heinrich2017} and \cite{heinrich2018}, artificially increasing the significance of some hints of a high-redshift reionization component. 

Even so, the PCA analysis can still be quite useful in providing a simple and convenient approximate likelihood as described in \cite{heinrich2017}. One can then easily project different physical models onto the principal components to obtain constraints on the physical model parameters, effectively applying the priors of the physical model and alleviating some of the problems that arise when attempting to interpret the PCA results on their own. 

We have also described the \FlexKnot model, which we argue is better to use in a completely generic reionization reconstruction setting since physicality can be enforced exactly and the question of how many knots to use can be self-consistently dealt with via a Bayesian evidence computation. 

Using these models, we have presented the first generic reconstruction results from the \Planck 2016 intermediate \texttt{simlow} likelihood. We find no evidence for early reionization, instead only very tight upper limits on any contribution at $z\,{\gtrsim}\,15$. This is true even when using the as-argued sub-optimal PCA model, thus the qualitative conclusion of preference only for a late and fast reionization is quite robust to modeling choices. These results negate the need for explanations of an early reionization contribution, for example from metal-free stars \citep{miranda2017}. 

Recently, the EDGES collaboration has detected an absorption feature in the sky-averaged spectrum, presumably due to 21cm absorption by neutral hydrogen during the epoch of reionization \citep{bowman2018}. In the standard picture, the upper and lower frequencies of the feature indicate that photons from early stars were numerous enough to equilibrate the gas temperature and 21cm spin temperature near $z\,{=}\,20$, and then raise the spin temperature above the CMB temperature near $z\,{=}\,15$. Due to the amplitude and shape of the detected feature, however, some non-standard explanation is required. The results discussed here can limit such possible explanations, as no modifications to the standard picture must be made which would lead to reionizing the universe to a level large enough to violate the bound on the optical depth at $z\,{>}\,15$ given in Eq.~\eqref{eq:tauearly_simlow}. For example, \cite{ewall-wice2018} discuss an explanation of the EDGES signal in which a radio background from accretion onto the black hole remnants of metal-free stars causes the absorption feature to appear deeper against the continuum than in the standard scenario. It is shown that X-rays also produced by the accretion would result in some significant reionization at $z\,{>}\,15$, already in some tension with even the high $\tau(15,30)$ obtained by \cite{heinrich2018}; given the tighter upper bounds we find here, this explanation is now heavy disfavored in its simplest form. 

Regardless of the generic reconstruction models we consider, we find there is some sensitivity to whether a flat prior on $\tau$ is taken, or one which is flat on PCA mode amplitudes or knot positions and/or amplitudes. Other priors for this problem exist as well, for example Jeffreys priors or reference priors \citep{jeffreys1946,bernardo1979}, and these would be well worth exploring too. Nevertheless, it is unclear that a very strong argument could be made for which particular prior is the correct one. Thus here we instead take the shifts in $\tau$ we find when switching between the priors we have considered as an unavoidable modeling uncertainty arising from the generic reconstruction. 

We showed that this uncertainty can have significant impact on determination of the sum of neutrino masses from future experiments. For example, different priors can cause a shift of $0.35\,\sigma$ on $\Sigma m_\nu$ from future \Planck + CMB-S4 + DESI-BAO measurements. One avenue for improvement would be to obtain something approaching a cosmic-variance-limited measurement of large-scale CMB polarization. Some other avenues exist as well.

For example, several direct constraints exist on the ionization state of the universe at various redshifts \citep[summarized, e.g., by][]{bouwens2015}. Here we have applied only one of them, mainly the stringent bound on the end of reionization from observations of the Gunn-Peterson optical depth in high-redshift quasars. We have done so in a fairly unsophisticated way by simply requiring that $x_{\rm e}(6)\,{=}\,x_{\rm e}^{\rm max}$. A more detailed study of the impact that these direct constraints have is warranted.

Another approach is to use physically motivated models of reionization, rather than the generic {\it unphysical} reconstruction methods presented here. These could potentially shrink the prior parameter space further to where priors become unimportant. The challenge here is finding models which have parameters that can be marginalized over to accurately represent uncertainty as to the physics of reionization and the types of sources which can contribute ionization radiation. In addition, with physical models, we may better be able to fold in constraints from other sources, such as measurements of the patchy kinetic Sunyaev-Zeldovich effect which can further constrain reionization \citep{gruzinov1998, knox1998, zahn2012, planck2014-a25}. We have not applied these bounds here as it is not straightforward how to do so for the generic reionization histories we consider, but in principle these could reduce our dependence on priors. Additionally, using novel analysis methods \citep{smith2016,ferraro2018}, experiments like CMB-S4 could potentially measure this effect to high significance. Further down the road, methods described in \cite{liu2016a} or \cite{meyers2017} could also push beyond the CMB cosmic variance limit we have calculated and remove the uncertainty on other cosmological parameters due to $\tau$ entirely. Future prospects are thus optimistic, although work remains to be done.

\begin{acknowledgements}
We thank Zhen Pan for his forecasting code which we adapted to use in this work, Torsten Ensslin, Douglas Scott, and Martin White for their helpful comments on the draft of this paper, and Silvia Galli, William Handley, and Wayne Hu for useful discussions. MM was supported by the Labex ILP (reference ANR-10-LABX-63). This work was aided by the use of several software packages, including CAMB \citep{lewis2000}, CosmoSlik \citep{millea2017b}, getdist \citep{antonylewis2016}, Julia \citep{bezanson2017}, and PolyChord \citep{handley2015a,handley2015}.
\end{acknowledgements}

\appendix

\section{Maximum entropy prior flattening}
\label{app:flattau}

Here we will show how to construct the maximum entropy prior used in Sect.~\ref{sec:flatten_tau}. This is the maximally uninformative prior on the input parameters of our model, $(\theta_1, \theta_2, ...)$, for which the induced prior on $\tau=\tau(\theta_1, \theta_2, ...)$ is flat. Here, the $\theta_i$ can represent the PCA model amplitudes, the knot positions and/or amplitudes, or any other parameters. 

We first begin with a simplified example: what is the maximum entropy prior on two parameters, $a$ and $b$, each with support on $0<a,b<1$, for which the prior on the sum, $a+b$, is flat between $0<a+b<2$? In this analogy, $a$ and $b$ are the $\theta_i$ parameters, $a+b$ is like $\tau$, and the support on [0,1] is the physicality region.

The entropy of the probability distribution $p(a,b)$ is
\begin{align}
H[p(a,b)] = - \int_0^1 da \int_0^1 db   \, p(a,b) \ln p(a,b).
\end{align}
We can express the problem of finding the $p(a,b)$ which maximizes $H$, subject to some constraints, as a Lagrange multiplier problem. Our constraint is that the transformed probability distribution for the sum, $p(a+b)$, is flat. We will write this constraint in what may seem as an odd form, but which is conducive to solving the Lagrange multiplier problem. We will require that the moments of $p(a,b)$ with respect to $a+b$ are those of a flat distribution between 0 and 2. This is to say, that $\langle a+b \rangle\,{=}\,1$, $\langle (a+b)^2 \rangle\,{=}\,8/3$, etc... By specifying an infinite number of moments, we guarantee our target distribution $p(a+b)$ is exactly flat. Our constraints are thus that
\begin{align}
\label{eq:priorcons}
\int_0^1 da \int_0^1 db \, (a+b)^n p(a,b) = c_n ,
\end{align}
with
\begin{align}
c_n = \frac{1}{2} \int_0^2 dx \, x^n
\end{align}
for all $n\,{=}0...\infty$. Note that for $n\,{=}\,0$ we simply have that the integral over $p(a,b)$ is unity, which guarantees that it is a probability distribution. Higher moments fix $p(a+b)$ to be flat as desired. As a Lagrange multiplier problem, we are maximizing the functional, 
\begin{align}
F[p(a,b)] =  &\int_0^1 da \int_0^1 db \;\Biggl\{ -p(a,b) \ln p(a,b) \nonumber \\ &+   \sum_{i=0}^\infty \lambda_i \left[(a+b)^n p(a,b) - c_n \right]\Biggr\},
\end{align}
where the $\lambda_i$ are the Lagrange multipliers. 
Setting the variation of $F$ with respect to $p(a,b)$ to zero gives us that
\begin{align}
\label{eq:elsoln}    
p(a,b) = \exp \left[-1+\sum_{i=0}^\infty \lambda_i (a+b)^n\right].
\end{align}
Substituting this into each of the Eq.~\eqref{eq:priorcons} could then be used to solve for all of the $\lambda_i$, giving us the unique maximum entropy solution for $p(a,b)$. Rather than preforming this process explicitly, we will postulate the answer and show that it is the solution. Consider the probability distribution
\begin{align}
\label{eq:soln}
p(a,b) = 1/q(a+b),
\end{align}
where $q(a+b)$ is the transformed probability distribution for $a+b$ given the initial flat priors on $a$ and $b$,
\begin{align}
q(a+b) = \int_0^1 da' \int_0^1 db' \delta((a+b)-(a'+b')).
\end{align}
It is straightforward to substitute this into Eq.~\eqref{eq:soln} and then show that $p(a+b)$ is indeed flat between 0 and 2. This means it satisfies the infinite number of constraint equations. It remains to check that the variation of $F$ around this function is zero, which is tantamount to showing that it can be written in the form of Eq.~\eqref{eq:elsoln}. Since Eq.~\eqref{eq:elsoln} says that the log of our function is a Maclaurin series in $a+b$ and since our $p(a,b)$ is an elementary function of only $a+b$, this is also true. Therefore, we have shown Eq.~\eqref{eq:elsoln} is the unique maximum entropy probability distribution of $a,b$ for which the transformed distribution of $a+b$ is flat. 

The above proof trivially generalizes to $N$ variables and to any initial support by simply adding in more integrals and changing the limits to something other than 0 and 1. Similarly, one can easily replace $a+b$ with any desired function for a derived parameter. This proof thus applies to the case of flattening the $\tau$ prior discussed in Sec.~\ref{sec:flatten_tau}. For a more rigorous and general proof, see \cite{handley2018}.

\bibliographystyle{aa} 
\bibliography{remnu,Planck_bib}

\begin{thebibliography}{53}
\expandafter\ifx\csname natexlab\endcsname\relax\def\natexlab#1{#1}\fi

\bibitem[{Abazajian {et~al.}(2016)Abazajian, Adshead, Ahmed, Allen, Alonso,
  Arnold, Baccigalupi, Bartlett, Battaglia, Benson, Bischoff, Borrill, Buza,
  Calabrese, Caldwell, Carlstrom, Chang, Crawford, Cyr-Racine, De~Bernardis,
  {de Haan}, {di Serego Alighieri}, Dunkley, Dvorkin, Errard, Fabbian, Feeney,
  Ferraro, Filippini, Flauger, Fuller, Gluscevic, Green, Grin, Grohs, Henning,
  Hill, Hlozek, Holder, Holzapfel, Hu, Huffenberger, Keskitalo, Knox, Kosowsky,
  Kovac, Kovetz, Kuo, Kusaka, Le~Jeune, Lee, Lilley, Loverde, Madhavacheril,
  Mantz, Marsh, McMahon, Meerburg, Meyers, Miller, Munoz, Nguyen, Niemack,
  Peloso, Peloton, Pogosian, Pryke, Raveri, Reichardt, Rocha, Rotti, Schaan,
  Schmittfull, Scott, Sehgal, Shandera, Sherwin, Smith, Sorbo, Starkman, Story,
  {van Engelen}, Vieira, Watson, Whitehorn, \& Kimmy~Wu}]{abazajian2016a}
Abazajian, K.~N., Adshead, P., Ahmed, Z., {et~al.} 2016, ArXiv e-prints, 1610,
  arXiv:1610.02743

\bibitem[{Abazajian {et~al.}(2015)Abazajian, Arnold, Austermann, Benson,
  Bischoff, Bock, Bond, Borrill, Calabrese, Carlstrom, Carvalho, Chang, Chiang,
  Church, Cooray, Crawford, Dawson, Das, Devlin, Dobbs, Dodelson, Dor{\'e},
  Dunkley, Errard, Fraisse, Gallicchio, Halverson, Hanany, Hildebrandt, Hincks,
  Hlozek, Holder, Holzapfel, Honscheid, Hu, Hubmayr, Irwin, Jones,
  Kamionkowski, Keating, Keisler, Knox, Komatsu, Kovac, Kuo, Lawrence, Lee,
  Leitch, Linder, Lubin, McMahon, Miller, Newburgh, Niemack, Nguyen, Nguyen,
  Page, Pryke, Reichardt, Ruhl, Sehgal, Seljak, Sievers, Silverstein, Slosar,
  Smith, Spergel, Staggs, Stark, Stompor, Vieregg, Wang, Watson, Wollack, Wu,
  Yoon, \& Zahn}]{abazajian2015}
Abazajian, K.~N., Arnold, K., Austermann, J., {et~al.} 2015, Astroparticle
  Physics, 63, 66

\bibitem[{Allison {et~al.}(2015)Allison, Caucal, Calabrese, Dunkley, \&
  Louis}]{allison2015}
Allison, R., Caucal, P., Calabrese, E., Dunkley, J., \& Louis, T. 2015,
  Physical Review D, 92

\bibitem[{Becker {et~al.}(2001)Becker, Fan, White, Strauss, Narayanan, Lupton,
  Gunn, Annis, Bahcall, Brinkmann, Connolly, Csabai, Czarapata, Doi, Heckman,
  Hennessy, Ivezi{\'c}, Knapp, Lamb, McKay, Munn, Nash, Nichol, Pier, Richards,
  Schneider, Stoughton, Szalay, Thakar, \& York}]{becker2001}
Becker, R.~H., Fan, X., White, R.~L., {et~al.} 2001, The Astronomical Journal,
  122, 2850

\bibitem[{Bernardo(1979)}]{bernardo1979}
Bernardo, J.~M. 1979, Journal of the Royal Statistical Society. Series B
  (Methodological), 41, 113

\bibitem[{Bezanson {et~al.}(2017)Bezanson, Edelman, Karpinski, \&
  Shah}]{bezanson2017}
Bezanson, J., Edelman, A., Karpinski, S., \& Shah, V. 2017, SIAM Review, 59, 65

\bibitem[{Bouwens {et~al.}(2015)Bouwens, Illingworth, Oesch, Caruana, Holwerda,
  Smit, \& Wilkins}]{bouwens2015}
Bouwens, R.~J., Illingworth, G.~D., Oesch, P.~A., {et~al.} 2015, The
  Astrophysical Journal, 811, 140

\bibitem[{Bowman {et~al.}(2018)Bowman, Rogers, Monsalve, Mozdzen, \&
  Mahesh}]{bowman2018}
Bowman, J.~D., Rogers, A. E.~E., Monsalve, R.~A., Mozdzen, T.~J., \& Mahesh, N.
  2018, Nature, 555, 67

\bibitem[{Colombo \& Pierpaoli(2009)}]{colombo2009}
Colombo, L. P.~L. \& Pierpaoli, E. 2009, New Astronomy, 14, 269

\bibitem[{Dodelson(2003)}]{dodelson2003}
Dodelson, S. 2003, Modern Cosmology

\bibitem[{Ewall-Wice {et~al.}(2018)Ewall-Wice, Chang, Lazio, Dor{\'e},
  Seiffert, \& Monsalve}]{ewall-wice2018}
Ewall-Wice, A., Chang, T.-C., Lazio, J., {et~al.} 2018, ArXiv e-prints, 1803,
  arXiv:1803.01815

\bibitem[{Fan {et~al.}(2002)Fan, Narayanan, Strauss, White, Becker, Pentericci,
  \& Rix}]{fan2002}
Fan, X., Narayanan, V.~K., Strauss, M.~A., {et~al.} 2002, The Astronomical
  Journal, 123, 1247

\bibitem[{Ferraro \& Smith(2018)}]{ferraro2018}
Ferraro, S. \& Smith, K.~M. 2018, ArXiv e-prints, 1803, arXiv:1803.07036

\bibitem[{Font-Ribera {et~al.}(2014)Font-Ribera, McDonald, Mostek, Reid, Seo,
  \& Slosar}]{font-ribera2014}
Font-Ribera, A., McDonald, P., Mostek, N., {et~al.} 2014, Journal of Cosmology
  and Astroparticle Physics, 2014, 023

\bibitem[{Gruzinov \& Hu(1998)}]{gruzinov1998}
Gruzinov, A. \& Hu, W. 1998, The Astrophysical Journal, 508, 435

\bibitem[{Hamimeche \& Lewis(2008)}]{hamimeche2008}
Hamimeche, S. \& Lewis, A. 2008, Physical Review D, 77, 103013

\bibitem[{Handley \& Millea(2018)}]{handley2018}
Handley, W. \& Millea, M. 2018, ArXiv e-prints, 1804, arXiv:1804.08143

\bibitem[{Handley {et~al.}(2015{\natexlab{a}})Handley, Hobson, \&
  Lasenby}]{handley2015a}
Handley, W.~J., Hobson, M.~P., \& Lasenby, A.~N. 2015{\natexlab{a}}, Monthly
  Notices of the Royal Astronomical Society: Letters, 450, L61

\bibitem[{Handley {et~al.}(2015{\natexlab{b}})Handley, Hobson, \&
  Lasenby}]{handley2015}
Handley, W.~J., Hobson, M.~P., \& Lasenby, A.~N. 2015{\natexlab{b}}, Monthly
  Notices of the Royal Astronomical Society, 453, 4385

\bibitem[{Hazra \& Smoot(2017)}]{hazra2017}
Hazra, D.~K. \& Smoot, G.~F. 2017, Journal of Cosmology and Astro-Particle
  Physics, 11, 028

\bibitem[{Heinrich \& Hu(2018)}]{heinrich2018}
Heinrich, C. \& Hu, W. 2018, ArXiv e-prints, 1802, arXiv:1802.00791

\bibitem[{Heinrich {et~al.}(2017)Heinrich, Miranda, \& Hu}]{heinrich2017}
Heinrich, C.~H., Miranda, V., \& Hu, W. 2017, Physical Review D, 95, 023513

\bibitem[{Hu \& Holder(2003)}]{hu2003}
Hu, W. \& Holder, G.~P. 2003, Physical Review D, 68, 023001

\bibitem[{Jeffreys(1946)}]{jeffreys1946}
Jeffreys, H. 1946, Proceedings of the Royal Society of London. Series A,
  Mathematical and Physical Sciences, 186, 453

\bibitem[{Knox {et~al.}(1998)Knox, Scoccimarro, \& Dodelson}]{knox1998}
Knox, L., Scoccimarro, R., \& Dodelson, S. 1998, Physical Review Letters, 81,
  2004

\bibitem[{Kogut {et~al.}(2011)Kogut, Fixsen, Chuss, Dotson, Dwek, Halpern,
  Hinshaw, Meyer, Moseley, Seiffert, Spergel, \& Wollack}]{kogut2011}
Kogut, A., Fixsen, D.~J., Chuss, D.~T., {et~al.} 2011, Journal of Cosmology and
  Astro-Particle Physics, 07, 025

\bibitem[{Levi {et~al.}(2013)Levi, Bebek, Beers, Blum, Cahn, Eisenstein,
  Flaugher, Honscheid, Kron, Lahav, McDonald, Roe, Schlegel, \& {representing
  the DESI collaboration}}]{levi2013}
Levi, M., Bebek, C., Beers, T., {et~al.} 2013, ArXiv e-prints, 1308,
  arXiv:1308.0847

\bibitem[{Lewis {et~al.}(2000)Lewis, Challinor, \& Lasenby}]{lewis2000}
Lewis, A., Challinor, A., \& Lasenby, A. 2000, The Astrophysical Journal, 538,
  473

\bibitem[{Lewis {et~al.}(2006)Lewis, Weller, \& Battye}]{lewis2006}
Lewis, A., Weller, J., \& Battye, R. 2006, Monthly Notices of the Royal
  Astronomical Society, 373, 561

\bibitem[{Lewis \& Xu(2016)}]{antonylewis2016}
Lewis, A. \& Xu, Y. 2016, Cmbant/Getdist: 0.2.6, Tech. rep., {Zenodo}

\bibitem[{Li(2011)}]{li2011}
Li, S. 2011, Asian Journal of Mathematics \& Statistics, 4, 66

\bibitem[{Liu {et~al.}(2016)Liu, Pritchard, Allison, Parsons, Seljak, \&
  Sherwin}]{liu2016a}
Liu, A., Pritchard, J.~R., Allison, R., {et~al.} 2016, Physical Review D, 93,
  043013

\bibitem[{Matsumura {et~al.}(2016)Matsumura, Akiba, Arnold, Borrill, Chendra,
  Chinone, Cukierman, de~Haan, Dobbs, Dominjon, Elleflot, Errard, Fujino, Fuke,
  Goeckner-wald, Halverson, Harvey, Hasegawa, Hattori, Hattori, Hazumi, Hill,
  Hilton, Holzapfel, Hori, Hubmayr, Ichiki, Inatani, Inoue, Inoue, Irie, Irwin,
  Ishino, Ishitsuka, Jeong, Karatsu, Kashima, Katayama, Kawano, Keating,
  Kibayashi, Kibe, Kida, Kimura, Kimura, Kohri, Komatsu, Kuo, Kuromiya, Kusaka,
  Lee, Linder, Matsuhara, Matsuoka, Matsuura, Mima, Mitsuda, Mizukami, Morii,
  Morishima, Nagai, Nagasaki, Nagata, Nakajima, Nakamura, Namikawa, Naruse,
  Natsume, Nishibori, Nishijo, Nishino, Nitta, Noda, Noguchi, Ogawa, Oguri,
  Ohta, Otani, Okada, Okamoto, Okamoto, Okamura, Rebeiz, Richards, Sakai, Sato,
  Sato, Segawa, Sekiguchi, Sekimoto, Sekine, Seljak, Sherwin, Shinozaki, Shu,
  Stompor, Sugai, Sugita, Suzuki, Suzuki, Tajima, Takada, Takakura, Takano,
  Takei, Tomaru, Tomita, Turin, Utsunomiya, Uzawa, Wada, Watanabe, Westbrook,
  Whitehorn, Yamada, Yamasaki, Yamashita, Yoshida, Yoshida, \&
  Yotsumoto}]{matsumura2016}
Matsumura, T., Akiba, Y., Arnold, K., {et~al.} 2016, Journal of Low Temperature
  Physics, 184, 824

\bibitem[{Meyers {et~al.}(2017)Meyers, Meerburg, {van Engelen}, \&
  Battaglia}]{meyers2017}
Meyers, J., Meerburg, P.~D., {van Engelen}, A., \& Battaglia, N. 2017, ArXiv
  e-prints, 1710, arXiv:1710.01708

\bibitem[{Millea(2017)}]{millea2017b}
Millea, M. 2017, Astrophysics Source Code Library, ascl:1701.004

\bibitem[{Miranda {et~al.}(2017)Miranda, Lidz, Heinrich, \& Hu}]{miranda2017}
Miranda, V., Lidz, A., Heinrich, C.~H., \& Hu, W. 2017, Monthly Notices of the
  Royal Astronomical Society, 467, 4050

\bibitem[{Mortonson \& Hu(2008{\natexlab{a}})}]{mortonson2008}
Mortonson, M.~J. \& Hu, W. 2008{\natexlab{a}}, The Astrophysical Journal, 672,
  737

\bibitem[{Mortonson \& Hu(2008{\natexlab{b}})}]{mortonson2008a}
Mortonson, M.~J. \& Hu, W. 2008{\natexlab{b}}, The Astrophysical Journal, 686,
  L53

\bibitem[{Obied {et~al.}(2018)Obied, Dvorkin, Heinrich, Hu, \&
  Miranda}]{obied2018}
Obied, G., Dvorkin, C., Heinrich, C., Hu, W., \& Miranda, V. 2018, ArXiv
  e-prints, 1803, arXiv:1803.01858

\bibitem[{Okamoto \& Hu(2003)}]{okamoto2003}
Okamoto, T. \& Hu, W. 2003, Physical Review D, 67, 083002

\bibitem[{Pan \& Knox(2015)}]{pan2015}
Pan, Z. \& Knox, L. 2015, Monthly Notices of the Royal Astronomical Society,
  454, 3200

\bibitem[{{Planck Collaboration}(2005)}]{planck2005-bluebook}
{Planck Collaboration}. 2005, ESA publication ESA-SCI(2005)/01
  [\eprint[arXiv]{astro-ph/0604069}]

\bibitem[{{\sorthelp{Planck Collaboration 2015B}}{Planck Collaboration
  II}(2016)}]{planck2014-a03}
{\sorthelp{Planck Collaboration 2015B}}{Planck Collaboration II}. 2016, \aap,
  594, A2

\bibitem[{{\sorthelp{Planck Collaboration 2015T}}{Planck Collaboration
  XX}(2016)}]{planck2014-a24}
{\sorthelp{Planck Collaboration 2015T}}{Planck Collaboration XX}. 2016, \aap,
  594, A20

\bibitem[{{\sorthelp{Planck Collaboration IntZU}}{Planck Collaboration Int.
  XLVI}(2016)}]{planck2014-a10}
{\sorthelp{Planck Collaboration IntZU}}{Planck Collaboration Int. XLVI}. 2016,
  \aap, 596, A107

\bibitem[{{\sorthelp{Planck Collaboration IntZV}}{Planck Collaboration Int.
  XLVII}(2016)}]{planck2014-a25}
{\sorthelp{Planck Collaboration IntZV}}{Planck Collaboration Int. XLVII}. 2016,
  \aap, 596, A108

\bibitem[{{\sorthelp{Planck Collaboration IntZZA}}{Planck Collaboration
  LI}(2017)}]{planck2016-LI}
{\sorthelp{Planck Collaboration IntZZA}}{Planck Collaboration LI}. 2017, \aap,
  607, A95

\bibitem[{Smith \& Ferraro(2016)}]{smith2016}
Smith, K.~M. \& Ferraro, S. 2016, ArXiv e-prints, 1607, arXiv:1607.01769

\bibitem[{Smith {et~al.}(2012)Smith, Hanson, LoVerde, Hirata, \&
  Zahn}]{smith2012}
Smith, K.~M., Hanson, D., LoVerde, M., Hirata, C.~M., \& Zahn, O. 2012, Journal
  of Cosmology and Astroparticle Physics, 2012, 014

\bibitem[{{The COrE Collaboration} {et~al.}(2011){The COrE Collaboration},
  Armitage-Caplan, Avillez, Barbosa, Banday, Bartolo, Battye, Bernard, {de
  Bernardis}, Basak, Bersanelli, Bielewicz, Bonaldi, Bucher, Bouchet,
  Boulanger, Burigana, Camus, Challinor, Chongchitnan, Clements, Colafrancesco,
  Delabrouille, De~Petris, De~Zotti, Dickinson, Dunkley, Ensslin, Fergusson,
  Ferreira, Ferriere, Finelli, Galli, Garcia-Bellido, Gauthier, Haverkorn,
  Hindmarsh, Jaffe, Kunz, Lesgourgues, Liddle, Liguori, Lopez-Caniego, Maffei,
  Marchegiani, Martinez-Gonzalez, Masi, Mauskopf, Matarrese, Melchiorri,
  Mukherjee, Nati, Natoli, Negrello, Pagano, Paoletti, Peacocke, Peiris,
  Perroto, Piacentini, Piat, Piccirillo, Pisano, Ponthieu, Rath, Ricciardi,
  Rubino~Martin, Salatino, Shellard, Stompor, Urrestilla, Van~Tent, Verde,
  Wandelt, \& Withington}]{thecorecollaboration2011}
{The COrE Collaboration}, Armitage-Caplan, C., Avillez, M., {et~al.} 2011,
  ArXiv e-prints, 1102, arXiv:1102.2181

\bibitem[{V{\'a}zquez {et~al.}(2012)V{\'a}zquez, Bridges, Hobson, \&
  Lasenby}]{vazquez2012}
V{\'a}zquez, J.~A., Bridges, M., Hobson, M.~P., \& Lasenby, A.~N. 2012, Journal
  of Cosmology and Astro-Particle Physics, 06, 006

\bibitem[{Villanueva-Domingo {et~al.}(2018)Villanueva-Domingo, Gariazzo,
  Gnedin, \& Mena}]{villanueva-domingo2018}
Villanueva-Domingo, P., Gariazzo, S., Gnedin, N.~Y., \& Mena, O. 2018, Journal
  of Cosmology and Astro-Particle Physics, 04, 024

\bibitem[{Zahn {et~al.}(2012)Zahn, Reichardt, Shaw, Lidz, Aird, Benson, Bleem,
  Carlstrom, Chang, Cho, Crawford, Crites, {de Haan}, Dobbs, Dore, Dudley,
  George, Halverson, Holder, Holzapfel, Hoover, Hou, Hrubes, Joy, Keisler,
  Knox, Lee, Leitch, Lueker, Luong-Van, McMahon, Mehl, Meyer, Millea, Mohr,
  Montroy, Natoli, Padin, Plagge, Pryke, Ruhl, Schaffer, Shirokoff, Spieler,
  Staniszewski, Stark, Story, {van Engelen}, Vanderlinde, Vieira, \&
  Williamson}]{zahn2012}
Zahn, O., Reichardt, C.~L., Shaw, L., {et~al.} 2012, The Astrophysical Journal,
  756, 65

\end{thebibliography}

\end{document}